\documentclass[twocolumn,aps,prd]{revtex4-1}
\usepackage{amsmath,graphicx,bm}
\begin{document}

\def\spose#1{\hbox to 0pt{#1\hss}}
\def\simlt{\mathrel{\spose{\lower 3pt\hbox{$\mathchar"218$}}
    \raise 2.0pt\hbox{$\mathchar"13C$}}}
    
\title{Self-Similar Spherical Collapse with Tidal Torque}

\author{Phillip Zukin}
\email{zukin@mit.edu}
\author{Edmund Bertschinger}
\affiliation{Department of Physics, MIT, 77 Massachusetts Ave.,
Cambridge, MA 02139}

\begin{abstract}

N-body simulations have revealed a wealth of information about dark matter halos however their results are largely empirical. Using analytic means, we attempt to shed light on simulation results by generalizing the self-similar secondary infall model to include tidal torque. In this first of two papers, we describe our halo formation model and compare our results to empirical mass profiles inspired by N-body simulations. Each halo is determined by four parameters. One parameter sets the mass scale and the other three define how particles within a mass shell are torqued throughout evolution. We choose torque parameters motivated by tidal torque theory and N-body simulations and analytically calculate the structure of the halo in different radial regimes. We find that angular momentum plays an important role in determining the density profile at small radii. For cosmological initial conditions, the density profile on small scales is set by the time rate of change of the angular momentum of particles as well as the halo mass. On intermediate scales, however, $\rho\propto r^{-2}$, while $\rho\propto r^{-3}$ close to the virial radius. 
\end{abstract}

\pacs{04.50.Kd,04.50.-h,95.36.+x}

\maketitle

\section{Introduction}
\label{sec:intro}

The structure of dark matter halos affects our understanding of galaxy formation and evolution and has implications for dark matter detection. Progress in our understanding of dark matter halos has been made both numerically and analytically. Analytic treatments began with work by Gunn and Gott; they analyzed how bound mass shells that accrete onto an initially collapsed object can explain the morphology of the Coma cluster \cite{GunnGott72, Gott75} and elliptical galaxies \cite{Gunn77}. This continuous accretion process is known as secondary infall.  

Secondary infall introduces a characteristic length scale: the shell's turnaround radius $r_*$. This is the radius at which a particular mass shell first turns around. Since the average density is a decreasing function of distance from the collapsed object, mass shells initially farther away will turnaround later. This characteristic scale should be expected since the radius of a mass shell, like the radius of a shock wave in the Sedov Taylor solution, can only depend on the initial energy of the shell, the background density, and time \cite{Bert}. By imposing that the structure of the halo is self-similar -- all quantities describing the halo only depend on the background density, $r_{\rm{ta}}$ (the current turnaround radius), and lengths scaled to $r_{\rm{ta}}$ -- Bertschinger \cite{Bert} and Fillmore and Goldreich (hereafter referred to as FG) \cite{FG} were able to relate the asymptotic slope of the nonlinear density profile to the initial linear density perturbation.

Assuming purely radial orbits, FG analytically showed that the slope $\nu$ of the halo density distribution $\rho\propto r^{-\nu}$ falls in the range $2<\nu<2.25$ for $r/r_{\rm{ta}}\ll1$. This deviates strongly from N-body simulations which find $\nu\simlt1$ \cite{Aquarius,Graham} or $\nu\sim1.2$ \cite{Lactea} at their innermost resolved radius and observations of Low Surface Brightness and spiral galaxies which suggest $\nu\sim.2$ \cite{deBlok03} and the presence of cores \cite{Salucci1,Salucci2,Salucci3}. Though the treatment in FG assumes radial orbits whileΠorbits in simulations and observed galaxies contain tangential components, it is analytically tractable and does not suffer from resolution limits. Numerical dark matter simulations, on the other hand, do not make any simplifying assumptions and have finite dynamic range. Moreover, it is difficult to draw understanding from their analysis and computational resources limit the smallest resolvable radius, since smaller scales require more particles and smaller time steps.  It seems natural, then, to generalize the work done by FG in order to explain the features predicted in simulations and observed in galaxies. This paper, in particular, investigates how non-radial motion affects the structure of dark matter halos.

Numerous authors have investigated how angular momentum affects the asymptotic density profile. Ryden and Gunn analyzed the effects of non-radial motion caused by substructure \cite{RydenGunn} while others have examined how an angular momentum, or a distribution of angular momenta, assigned to each mass shell at turnaround, affects the structure of the halo \cite{Nusser,Hiotelis,WilliamsEtAl,Sikivie,DelPopolo,WhiteZaritsky,LeDelliou,Ascasibar}. Note that many of these authors do not impose self-similarity. Those that do assume that a shell's angular momentum remains constant after turnaround. 

This paper extends previous work by Nusser \cite{Nusser}. Assuming self-similarity, he analytically calculated the structure of the halo in different radial regimes for shells with constant angular momentum after turnaround. He found that the inclusion of angular momentum allows $0<\nu<2.25$. According to Hoffman and Shaham \cite{HoffmanShaham}, $\nu$ depends on the effective primordial power spectral index ($d \ln P/d \ln k$), which varies for different mass halos. For galactic size halos, Nusser's analytic work predicts $\nu\sim 1.3$, in disagreement with simulation results \cite{Aquarius,Lactea}. In order to address this discrepancy, we extend Nusser's work by including torque. We consistently keep track of a particle's angular momentum, allowing it to build up before turnaround because of tidal interactions with neighboring protogalaxies \cite{Hoyle} and to evolve after turnaround because of nonlinear effects within the halo. Moreover, we compare the predictions of our halo model to simulation results. 

Self-similar secondary infall requires $\Omega_m=1$ since a nonvanishing $\Omega_{\Lambda}$ introduces an additional scale. Applying self-similarity to halo formation in the $\Lambda$CDM model therefore requires approximations and a mapping to halos in an Einstein de-Sitter universe. We assume that the linear power spectrum and background matter density $\bar{\rho}_m$ today are equal in both universes so that the statistics, masses, and length scales of halos found in the two models are equivalent. Since the scale factor evolves differently in both universes, the halo assembly histories will differ.

In section \ref{sec:SS}, we define our self-similar system and torqueing parameters. In section \ref{sec:BT}, we set initial conditions and evolve the mass shells before turnaround. In section \ref{sec:AT}, we describe evolution after turnaround and then analyze the asymptotic behavior of the density profile at different scales in section \ref{sec:AB}. In section \ref{sec:structure}, we give numerical results, discuss the overall structure of the halo, and compare to N-body simulations. We conclude in section \ref{sec:disc}.    

\section{Self-Similar Definitions}
\label{sec:SS}

Here we explicitly define our self-similar system and derive constraints on the functional form of the mass distribution within a halo and the angular momentum of particles in a particular shell. 

If the infall process is self-similar, then the halo's appearance does not change once all lengths are scaled to the current turnaround radius. For our analysis, we define the current turnaround radius as $r_{\rm{ta}}(t)\equiv Ct^{\beta}$ where both $C$ and $\beta$ are positive constants. The exponent $\beta$, as we will find, depends on the initial perturbation spectrum. 

The evolution of a particular mass shell must depend on time $t$ and the shell's turnaround time $t_*$. More explicitly, assuming spherical symmetry, we have $r=R(t,t_*)$. We define a self-similar system as one in which every trajectory obeys the following scaling: 

\begin{equation} \label{SSDef}
R(\Lambda t,\Lambda t_*) = \Lambda^{\beta}R(t,t_*)
\end{equation}

\noindent where  $\Lambda$ is a constant. The above implies that the trajectory of one mass shell with turnaround time $t_1$ can be mapped to the trajectory of another mass shell with turnaround time $t_2 = \Lambda t_1$. The exponent $\beta$ follows since $R(t_1,t_1)/R(t_2,t_2)=(t_1/t_2)^{\beta}$. 

Each shell of a self-similar system must also follow the same equation of motion. From Newton's law, the radial equation of motion for a mass shell with angular momentum is given by:

\begin{equation} \label{SSEOM}
\ddot{R}(t,t_*)=-\frac{GM\Big(R(t,t_*),t\Big)}{R^2(t,t_*)}+\frac{L^2\Big(R(t,t_*),t,t_*\Big)}{R^3(t,t_*)}
\end{equation}

\noindent where dots denote derivatives with respect to the first argument, $M$ is the mass of the halo interior to $r$ and $L$ is the angular momentum per unit mass of a particle in the shell. Note that we enforce the mass to not depend explicitly on $t_*$, while the angular momentum can. As we will show below, this is physically motivated. From eq.\ (\ref{SSDef}), we find:
 
 \begin{equation} \label{SSdotDef}
 \ddot{R}(\Lambda t,\Lambda t_*)=\Lambda^{\beta-2}R(t,t_*)
 \end{equation}

\noindent Plugging in eqs.\ (\ref{SSDef}) and (\ref{SSdotDef}) into eq.\ (\ref{SSEOM}) and simplifying, we find:

\begin{eqnarray} \label{SSEOM2}
\ddot{R}(\Lambda t, \Lambda t_*) = &-&\Lambda^{3\beta-2}\frac{GM\Big(\Lambda^{-\beta}R(\Lambda t, \Lambda t_*),t\Big)}{R^2(\Lambda t, \Lambda t_*)} \nonumber\\
&+&\Lambda^{4\beta-2} \frac{L^2\Big(\Lambda^{-\beta}R(\Lambda t, \Lambda t_*),t,t_*\Big)}{R^3(\Lambda t, \Lambda t_*)}
\end{eqnarray}

\noindent Changing variables from $R(t,t_*)$ to $R(t,t_*)/Ct^{\beta}$ for the mass and angular momentum and rewriting eq.\ (\ref{SSEOM2}), we find:

\begin{eqnarray}
\ddot{R}(\Lambda t, \Lambda t_*) = &-&\Lambda^{3\beta-2}\frac{GM\Big(R(\Lambda t, \Lambda t_*)/C(\Lambda t)^{\beta},t\Big)}{R^2(\Lambda t, \Lambda t_*)} \nonumber\\
&+&\Lambda^{4\beta-2} \frac{L^2\Big(R(\Lambda t, \Lambda t_*)/C(\Lambda t)^{\beta},t,t_*\Big)}{R^3(\Lambda t, \Lambda t_*)} \nonumber \\
\end{eqnarray}

\noindent Relabeling coordinates and enforcing consistency with eq.\ (\ref{SSEOM}), we find the following constraints on the functional forms of the mass and angular momentum.

\begin{eqnarray}
M\Big(R(t,t_*)/Ct^{\beta},t\Big) &=& \Lambda^{3\beta-2}M\Big(R(t,t_*)/Ct^{\beta},t/\Lambda\Big) \\
L\Big(R(t,t_*),t,t_*\Big) &=& \Lambda^{2\beta-1}L\Big(R(t,t_*)/Ct^{\beta},t/\Lambda,t_*/\Lambda\Big) \nonumber \\
\end{eqnarray}

With the above in mind, we define the angular momentum per unit mass $L$ of a particle in a shell at $r$, and the density $\rho$ and mass $M$ of the halo as follows. 

\begin{eqnarray}
L(r,t)&=&B\frac{r_{\rm{ta}}^2(t)}{t}f(\lambda, t/t_*)\label{LDef} \\
\rho(r,t)&=&\rho_B(t)D(\lambda)\label{rhoDef} \\
M(r,t) &=& \frac{4\pi}{3}\rho_B(t)r_{\rm{ta}}^3(t){\cal{M}}(\lambda)\label{MDef}
\end{eqnarray} 

\noindent where $\lambda\equiv r/r_{\rm{ta}}(t)$ is the radius scaled to the current turnaround radius and $\rho_B=1/6\pi Gt^2$ is the background density for an Einstein de-Sitter (flat $\Omega_m=1$) universe.

Using eq.\ (\ref{SSDef}), it is straightforward to show:

\begin{equation} \label{lambdaDef}
\lambda(t,\Lambda t_*)=\lambda(t/\Lambda, t_*)
\end{equation}

\noindent Eq.\ (\ref{lambdaDef}) implies that if one can compute $\lambda(t,t_*)$ for a particular mass shell $t_*$ at all times, then one also knows the position of all other mass shells, labeled by $\Lambda t_*$ with varying $\Lambda$, at a particular time. This interpretation is very powerful and will be used later in order to calculate the mass profile after turnaround. 

If the mass profile ${\cal{M}(\lambda)}$ also depended explicitly on $t_*$, then the mass would not have to grow like the background mass enclosed in the current turnaround radius. This is clearly not physical. Hence we suppressed the explicit dependence on $t_*$. On the other hand, we've kept the dependence on $t_*$ in the angular momentum in order to have this extra freedom. Inspired by tidal torque theory and numerical simulations, in eq.\ (\ref{LDef}) we take $f$ to be:  

\begin{equation} \label{torque}
f(\lambda,t/t_*)= 
\begin{cases} 
\lambda^{-\gamma} & \text{if $t<t_*$}, \\ 
(t/t_*)^{\varpi+1-2\beta} & \text{if $t> t_*$}. 
\end{cases} 
\end{equation} 

\noindent The constant $B$ sets the amplitude of the angular momentum at turnaround while $\gamma$ ($\varpi)$ controls how quickly the angular momentum increases before (after) turnaround.  Constraints on $B$, $\gamma$, and $\varpi$ will be discussed in later sections.

We've assumed that the halo is spherically symmetric. While simulated halos are triaxial \cite{Hayashi}, the description above is meant to represent an average halo. Since there are no preferred directions in the universe, it should be expected that a statistically averaged halo is spherically symmetric. 

In the above, $L$ represents the angular momentum per unit mass of all particles in the shell. We impose that all particles in the shell have orbital planes that are randomly distributed. This implies that the total vector angular momentum of the mass shell, and hence the total angular momentum of the halo $\boldsymbol{J}$, vanishes. Hence, while individual particles on a mass shell gain angular momentum in random directions throughout evolution, on average the mass shell remains spherical. Therefore, like we've assumed above, only one radial equation of motion is necessary to describe the evolution of the shell.

Since our statistically averaged halo has a vanishing total angular momentum, this model cannot address the nonzero spin parameters observed in individually simulated halos \cite{Spin1,Spin2}. Nor can it reproduce the nonzero value of $\left\langle J^2\right\rangle$ expected from cosmological perturbation theory \cite{Peebles,White,Dorosh}. However, $\int L^2 dm$ where $dm$ is the mass of a shell, does not vanish for this model. We will use this quantity, which is a measure of the tangential dispersion in the halo, to constrain our torque parameters.
 
\section{Before Turnaround}
\label{sec:BT}

The trajectory of the mass shell after turnaround determines the halo mass profile. In order to start integrating at turnaround, however, the enclosed mass of the halo must be known. For the case of purely radial orbits, the enclosed mass at turnaround can be analytically calculated \cite{Bert,FG}.  For the case of orbits that have a time varying angular momentum, we must numerically evolve both the trajectory and ${\cal{M}}(\lambda)$ before turnaround in order to determine the enclosed mass at turnaround. 

The trajectory of a mass shell follows from Newton's law. We have:

\begin{equation}\label{EOM}
\frac{d^2r}{dt^2}=-\frac{GM(r,t)}{r^2}+\frac{L^2(r,t)}{r^3}
\end{equation}

\noindent Rewriting eq.\ (\ref{EOM}) in terms of $\lambda$ and $\xi\equiv \log(t/t_i)$, where $t_i$ is the initial time, and plugging in eqs.\ (\ref{LDef}), (\ref{MDef}) and (\ref{torque}), we find: 

\begin{equation}\label{EOMSS}
\frac{d^2\lambda}{d\xi^2}+(2\beta-1)\frac{d\lambda}{d\xi}+\beta(\beta-1)\lambda=-\frac{2}{9}\frac{{\cal{M}}(\lambda)}{\lambda^2}+B^2\lambda^{-2\gamma-3}
\end{equation}

\noindent The angular momentum before turnaround was chosen so that eq.\ (\ref{EOMSS}) does not explicitly depend on $\xi$. This allows for a cleaner perturbative analysis. Since $r$ is an approximate power law in $t$ at early times, we still have the freedom to choose a particular torque model inspired by tidal torque theory. This will be discussed at the end of this section.  

In order to numerically solve the above equation, one must know ${\cal{M}}(\lambda)$, a function we do not have a priori.  Before turnaround, however, the enclosed mass of a particular shell remains constant throughout evolution since no shells cross. Taking advantage of this, we relate $dr/dt$ to ${\cal{M}}$ by taking a total derivative of eq.\ (\ref{MDef}). We find:

\begin{equation}\label{drdt}
\left(\frac{dr}{dt}\right)_M=\beta\frac{r}{t}-(3\beta-2)Ct^{\beta-1}\frac{{\cal{M}}}{{\cal{M}}^{'}}
\end{equation}

\noindent In the above, a prime represents a derivative taken with respect to $\lambda$. Taking another derivative of the above with respect to time, plugging into eq.\ (\ref{EOM}) and simplifying, we find an evolution equation for ${\cal{M}}$:

\begin{eqnarray}\label{Mevolve}
\beta(\beta-1)\lambda+(3\beta-2)(\beta-1)\frac{{\cal{M}}}{{\cal{M}}^{'}}- \\(3\beta-2)^2\frac{{\cal{M}}^2{\cal{M}}^{''}}{({\cal{M}}^{'})^3} \nonumber
= -\frac{2}{9}\frac{\cal{M}}{\lambda^2}+B^2\lambda^{-2\gamma-3} \nonumber
\end{eqnarray}

Given eqns.\ (\ref{EOMSS}) and (\ref{Mevolve}), we must now specify initial conditions when $\lambda\gg1$. We assume the following perturbative solutions for $D(\lambda)$ and $\lambda(\xi)$ valid at early times. ${\cal{M}}(\lambda)$ follows from eq.\ (\ref{MDef}).

\begin{eqnarray}
D(\lambda)&=&1+\delta_1\lambda^{-n}+\delta_2\lambda^{-p}+... \label{Dinit} \\ 
{\cal{M}}(\lambda)&=&\lambda^3\left(1+\frac{3\delta_1}{3-n}\lambda^{-n}+\frac{3\delta_2}{3-p}\lambda^{-p}+...\right) \label{Minit} \\ 
\lambda(\xi)&=& \lambda_0e^{(2/3-\beta)\xi}(1+\lambda_1e^{\alpha_1\xi}+\lambda_2e^{\alpha_2\xi}+...) \label{LambdaInit}
\end{eqnarray} 

\begin{figure}[t]
  \begin{center}
    \includegraphics[height = 70mm, width=\columnwidth]{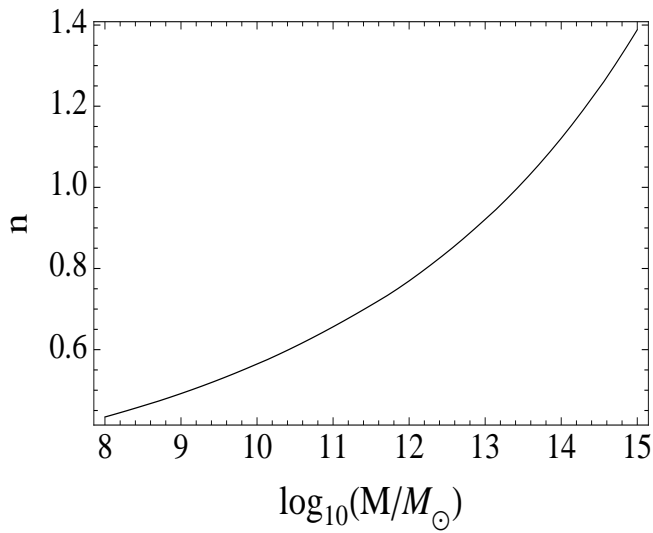}
  \end{center}
\caption{The variation of model parameter $n$ with halo mass. Larger mass halos map to steeper initial density profiles.} \label{fig:n}
\end{figure}

\noindent In the above, $n$ characterizes the first order correction to the background density. It is related to the FG parameter $\epsilon$ through $n=3\epsilon$. It is also related to the effective power spectral index $n_{\rm{eff}}=d \ln P/d\ln k$ through $n=n_{\rm{eff}}+3$ \cite{HoffmanShaham}. Since $n_{\rm{eff}}$ depends on scale and hence halo mass (Appendix \ref{sec:neff}), we have a relationship between $n$ and halo mass. As Figure\ \ref{fig:n} shows, larger mass halos have larger $n$. This is expected since larger smoothing lengths imply steeper initial density profiles. As in FG, we restrict $0<n<3$ so that the density decreases with radius while the mass increases. We examine this whole range for completeness even though $n>1.4$ corresponds to objects larger than galaxy clusters. The exponent $p$ characterizes the next order correction to the background density caused by angular momentum. Consistency with our perturbative expansion (eqs.\ \ref{Dinit} through \ref{LambdaInit}) demands that we take $n<p<2n$. However, it is straightforward to generalize to other cases (Section \ref{sec:TTT}). As we show below, the constants $\{\delta_2, \lambda_1,\lambda_2,\alpha_1,\alpha_2\}$ are set by the equations of motion. The constants $\{\delta_1,\lambda_0\}$ are set by boundary conditions. Plugging eq.\ (\ref{Minit}) into eq.\ (\ref{Mevolve}) and enforcing equality between terms proportional to $\lambda^{1-n}$ we find two possible solutions.

\begin{eqnarray}
\beta &=& \frac{2}{3}\left(1+\frac{1}{n}\right) \\
\beta &=& \frac{2}{3}\left(1-\frac{3}{2n}\right)
\end{eqnarray} 

\noindent These represent the two solutions to the second order differential equation (eq.\ \ref{Mevolve}). For the first case, the turnaround radius grows faster than the Hubble flow while for the second it grows slower. Hence, the first solution represents the growing mode of the perturbation while the second is the decaying mode. Since we are interested in the growth of halos, we will only consider the growing mode from now on. Next, imposing $p=2\gamma+4$ and enforcing equality between terms proportional to $\lambda^{1-p}$ in eq.\ (\ref{Mevolve}), we find:

\begin{equation}\label{delta2}
\delta_2=\frac{9n^2B^2(p-3)}{2(p-n)(3n+2p)}
\end{equation}

\noindent Comparing eqs.\ (\ref{Minit}) and (\ref{delta2}), we see that the correction to the initial mass caused by angular momentum is negative since $p>n$. This is expected since the angular momentum acts against gravity. 

Next, we find constraints for the parameters in eq.\ (\ref{LambdaInit}). Plugging in eq.\ (\ref{LambdaInit}) into eq.\ (\ref{EOMSS}) and setting terms linear in $\delta_1$, $\delta_2$, $\lambda_1$, and $\lambda_2$ equal to each other, we find:

\begin{eqnarray} 
\alpha_1&=&\frac{2}{3} \label{alpha1} \\
\lambda_1&=&\frac{\delta_1}{n-3}\lambda_0^{-n} \\
\alpha_2&=&p\left(\beta-\frac{2}{3}\right) \label{alpha2} \\
\lambda_2&=&\frac{\delta_2}{p-3}\lambda_0^{-p} \label{param}
\end{eqnarray}

\begin{figure}[t]
  \begin{center}
    \includegraphics[height = 70mm, width=\columnwidth]{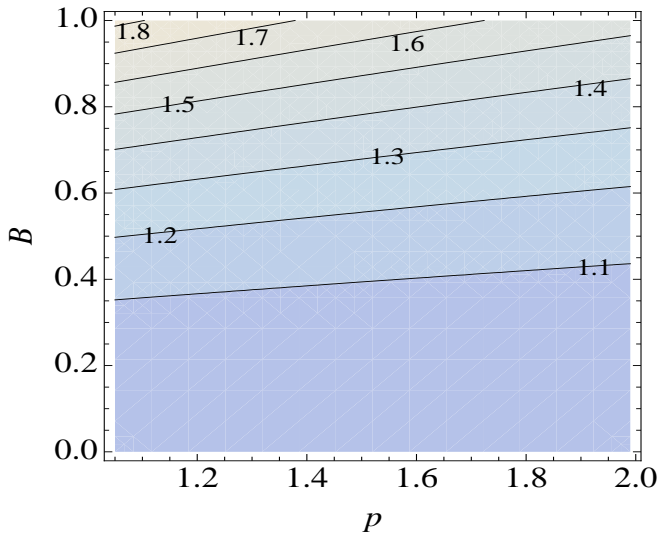}
  \end{center}
\caption{Contour plot of $16{\cal{M}}(1)/9\pi^2$ for $n=1$ as a function of torquing parameters $B$ and $p$. Smaller $p$ and larger $B$ result in larger torques on the mass shell. These larger torques require bigger enclosed masses at turnaround, in order to counteract the stronger angular momentum barrier.} \label{fig:contour}
\end{figure}

Eqns.\ (\ref{Dinit}) through (\ref{param}) set the initial conditions for eqs\ (\ref{EOMSS}) and (\ref{Mevolve}). We evolve $\lambda$ and ${\cal{M}}$ and choose $\{\lambda_0,\delta_1\}$ such that turnaround occurs ($d\lambda/d\xi=-\lambda\beta$) when $\lambda=1$. The free parameters for evolution before turnaround are $\{n,B,p\}$. For the zero angular momentum case analyzed in FG, the enclosed mass is the same regardless of $n$; ${\cal{M}}(1) = (3\pi/4)^2$. Including torque, however, we find that the enclosed mass depends on the parameters $B$ and $p$. Figure\ \ref{fig:contour} shows a contour plot of $16{\cal{M}}(1)/9\pi^2$ for $n=1$. As expected, the enclosed mass at turnaround must be larger than the no-torque case in order to overcome the additional angular momentum barrier. $B$ sets the amplitude of the angular momentum  while $p$ controls how the angular momentum grows in comparison to the mass perturbation. Larger $B$ and smaller $p$ correspond to stronger torques on the mass shell. Contour plots with different values of $n$ give the same features.
   
The above perturbative scheme ensures that the inclusion of angular momentum preserves cosmological initial conditions. Analyzing eq.\ (\ref{LambdaInit}) for $\xi\rightarrow-\infty$, and using eqs.\ (\ref{alpha1}) and (\ref{alpha2}), we see that since $p>n$,  the angular momentum correction is subdominant to the density perturbation correction. More importantly, at early times, the shell moves with the Hubble flow. Last, in order to be consistent with cosmological initial conditions, the angular momentum of particles within a mass shell must vanish at early times. Imposing $r\propto t^{2/3}$, and plugging into eq.\ (\ref{LDef}), we find that $L\propto t^{(1+p/n)/3}$. Hence, for all values of $p$ we consider, the angular momentum vanishes at early times and has a value of $Br^2_{*}/t_*$ at turnaround. Note that shells which turn around later have larger angular momentum. 

\subsection{Tidal Torque Theory}
\label{sec:TTT}

According to Hoyle's tidal torque mechanism, mass shells before turnaround gain angular momentum through their interactions with the tidal fields of neighboring protogalaxies \cite{Hoyle}. Peebles claimed that the angular momentum of protogalaxies in an Einstein de-Sitter universe grows as $t^{5/3}$ \cite{Peebles} while Doroshkevich showed that for non-spherical regions, the angular momentum grows as $t$ \cite{Dorosh,White}. White confirmed Doroshkevich's analysis with N-body simulations \cite{White}. Since the net angular momentum of our model's halo vanishes, we will instead use $\tilde{\sigma}^2$, defined below, to constrain $p$ and $B$. 

\begin{equation}\label{sigma}
\tilde{\sigma}^2\equiv \int_{V_L} dm |(\boldsymbol{r}-\boldsymbol{r}_0)\times(\boldsymbol{v}-\boldsymbol{v}_0)|^2
\end{equation}

\noindent In the above, we integrate over the Lagrangian volume $V_L$ of the halo, $\boldsymbol{r}$ and $\boldsymbol{v}$ are the physical radius and velocity of particles within the halo, $\boldsymbol{r}_0$ is the center of mass of the halo and $\boldsymbol{v}_0\equiv \boldsymbol v(\boldsymbol{r}_0)$. As described earlier in this section, $\lambda\gg1$ corresponds to early times when the halo is linear. Therefore, since our model represents a statistically averaged halo, we can calculate $\left\langle\tilde{\sigma}^2\right\rangle$ using cosmological linear perturbation theory and compare to expectations from our model. 

Using the Zel'dovich approximation \cite{zeldovich}, assuming a spherical Lagrangian volume with radius $R$, and working to first order, we find:

\begin{equation}\label{sigmaWhite}
\left\langle\tilde{\sigma}^2\right\rangle_M = 6a^4\dot{D}^2Mx^2_{\rm{max}}A^2(R)
\end{equation}

\noindent where $M$ is the mass of the halo, $a$ is the scale factor, $D$ is the linear growth factor, dots denote derivatives with respect to proper time, $x_{\rm{max}}$ is the lagrangian radius of the volume, $R$ is the spherical top hat radius for a halo of mass $M$ and $A(R)$ is a time independent function defined in Appendix \ref{sec:TT} which has units of length. Note that the scale factor $a$ and $\dot{D}$ are the only  quantities which vary with time. For a matter dominated universe, $(\left\langle\tilde{\sigma}^2\right\rangle_M)^{1/2}\propto t$, just as in the White analysis. This is expected since the Lagrangian mass is time independent.  

Next we calculate eq.\ (\ref{sigma}) from the perspective of our model. Using eqns.\ (\ref{LDef}), (\ref{MDef}), and (\ref{torque}) and assuming first order corrections to $M$ are negligible, we find:

\begin{eqnarray}\label{sigmaModel}
\tilde{\sigma}^2&=&\int^{r_{\rm{max}}}_{r_{\rm{min}}}B^2\frac{r^4_{ta}}{t^2}\lambda^{-2\gamma}\frac{\partial M(r,t)}{\partial r}dr \nonumber \\
&=& \frac{4\pi}{3-2\gamma}\frac{B^2\rho_B(t)r^7_{ta}(t)}{t^2}\left[\lambda^{3-2\gamma}_{\rm{max}}(t)-\lambda^{3-2\gamma}_{\rm{min}}(t)\right] \nonumber \\
\end{eqnarray}

\noindent The lower limit of integration sets an effective smoothing length which we choose to be $\lambda\gg1$ so that we only count shells that are still described by linear theory. The upper limit of integration is required since all the mass in the universe does not go into the halo. Since $p=2\gamma+4$ and $n<p<2n$, then for the range of $n$ we consider, $2\gamma<3$ and the angular momentum of the protogalaxy is dominated by shells close to $r_{\rm{max}}$. Equating eqs.\ (\ref{sigmaWhite}) and (\ref{sigmaModel}) and assuming the first order corrections to $r_{\rm{max}}$ in eq.\ (\ref{LambdaInit}) are negligible, we find $p=2n$ and: 

\begin{eqnarray}\label{pBconstraints}
B=\frac{2}{3}\sqrt{2(7-2n)}{\cal{M}}(1)^{(n-1)/3}\frac{A(R)}{R}
\end{eqnarray}

\noindent Eqs.\ (\ref{sigmaWhite}) and (\ref{pBconstraints}) are derived in Appendix \ref{sec:TT}. Note that the perturbative analysis, presented above, which is used to calculate  ${\cal{M}}(1)$ is not valid for $p=2n$. Redoing the analysis for this special case, we find:

\begin{eqnarray}\label{perturb2n}
\alpha_2&=& \frac{4}{3} \\
\delta_2&=& \frac{9}{14}B^2(2n-3)+\frac{(7n-17)(2n-3)}{7(n-3)^2}\delta^2_1 \\
\lambda_2&=&\frac{9}{14}\lambda^{-2n}_0\left(B^2-\frac{2}{3}\frac{\delta^2_1}{(n-3)^2}\right)
\end{eqnarray}

For the remainder of this paper we impose $p=2n$, so that angular momentum grows in accordance with cosmological perturbation theory, and set $n$ according to the halo mass. Unfortunately, when comparing to N-body simulations (Section \ref{nbody}), eq.\ (\ref{pBconstraints}) overestimates the angular momentum of particles at turnaround by a factor of 1.5 to 2.3. We discuss possible reasons for this discrepancy in Appendix \ref{sec:TT}. For convenience, $B_{1.5}$ ($B_{2.3}$) denotes $B$ calculated using eq.\ (\ref{pBconstraints}) with the right hand side divided by 1.5 (2.3). As described above, ${\cal{M}}(1)$ in eq.\ (\ref{pBconstraints}) depends on $B$. Therefore, in order to find $B$, we calculate $B$ and  ${\cal{M}}(1)$ iteratively until eq.\ (\ref{pBconstraints}) is satisfied. 

\begin{figure}[t]
  \begin{center}
    \includegraphics[height = 70mm, width=\columnwidth]{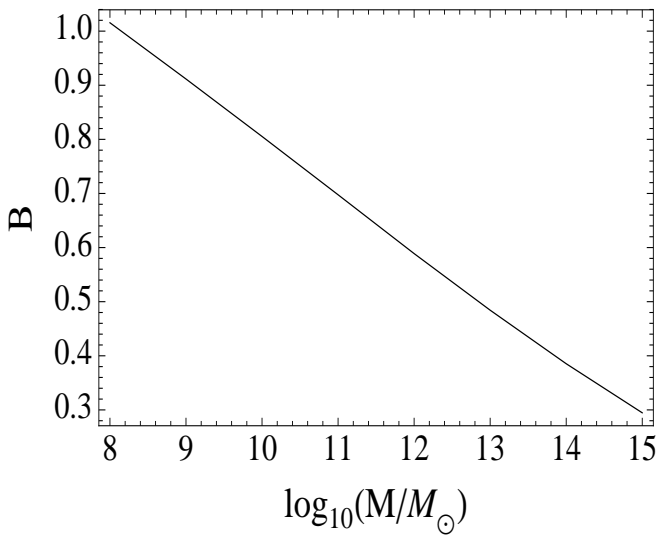}
  \end{center}
\caption{The variation of model parameter $B$ with halo mass. $B$ sets the angular momentum of particles at turnaround.} \label{fig:nB}
\end{figure}

The relationship between $n$ and $n_{\rm{eff}}$ as well as tidal torque theory implies that $\{n,p,B\}$ are all set by the halo mass. Figure\ \ref{fig:nB} shows the variation of $B$ with halo mass. $A(R)$ increases with halo mass since more power at large scales is included; $R$ also increases with halo mass. These two competing effects cause a slight variation in $B$ over seven orders of magnitude in halo mass.

\section{After Turnaround}
\label{sec:AT}

Given the enclosed mass found at turnaround, we now solve for the trajectory and mass profile after turnaround. For convenience, we redefine the time variable to be $\xi \equiv \ln(t/t_{ta})$, where $t_{ta}$ is the current turnaround time. The trajectory's evolution equation after turnaround, with the appropriate torque model (eq.\ \ref{torque}), is shown below.

\begin{equation}\label{EOMSS2}
\frac{d^2\lambda}{d\xi^2}+(2\beta-1)\frac{d\lambda}{d\xi}+\beta(\beta-1)\lambda=-\frac{2}{9}\frac{{\cal{M}}(\lambda)}{\lambda^2}+\frac{B^2}{\lambda^3}e^{2(\varpi+1-2\beta)\xi}
\end{equation}

The torque model after turnaround was chosen to not explicitly depend on $r$, since $r$ begins to oscillate on a much faster timescale than the growth of the halo. Nusser \cite{Nusser} and Sikivie et al. \cite{Sikivie} focused on the case $\varpi=0$. However, as was discussed before, this results in density profiles steeper than what is predicted by numerical simulations.

There  are a number of dynamical processes that can cause a particle's angular momentum to evolve after turnaround. Dynamical friction \cite{Chandra} transfers the angular momentum of massive bound objects -- like black holes, globular clusters and merging satellite galaxies -- to the background halo. A massive black hole at the center of the halo that dominates the potential at small scales tends to make the velocity dispersion isotropic \cite{Gerhard,Merritt,Cruz}. Bars \cite{Kalnajs,Dehnen} and supermassive black hole binaries \cite{Milos,Sesana} are also expected to perturb the dark matter velocity distribution. While the torque model proposed after turnaround is clearly very simplistic and may not accurately describe some of the above phenomena, it still allows us to get intuition for how torques acting on mass shells change the structure of the halo. 

Analytically calculating $\varpi$ is difficult since the halo after turnaround is nonlinear. In Appendix \ref{sec:ATE}, we show in a simplistic manner how $\varpi$ is sourced by substructure and argue that dark matter dominated substructure should cause steeper density profiles than baryon dominated substructure. In order to properly constrain $\varpi$, N-body simulations are required. This is beyond the scope of this work. 

The initial conditions for eq.\ (\ref{EOMSS2}), enforced in the above section, are $\lambda(\xi=0)=1$ and $d\lambda/d\xi(\xi=0)=-\beta$. As discussed before, self-similarity implies that all mass shells follow the same trajectory $\lambda(\xi)$. Hence, $\lambda(\xi)$ can either be interpreted as labeling the location of a particular mass shell at different times, or labeling the location of all mass shells at a particular time. We take advantage of the second interpretation in order to calculate the mass profile. 

After turnaround, shells cross since dark matter is collisionless. Therefore, the mass interior to a particular shell does not stay constant. However, since $\lambda(\xi)$ specifies the location of all mass shells at a particular time, the mass interior to a given scale is simply the sum of all mass shells interior to it. The mass profile is then given by \cite{Bert,FG}: 

\begin{eqnarray}\label{Mass}
{\cal{M}}(\lambda)&=&\frac{2}{n}{\cal{M}}(1)\int^{\infty}_0d\xi \exp[-(2/n)\xi] H[\lambda-\lambda(\xi)] \nonumber \\
&=&{\cal{M}}(1)\sum_i(-1)^{i-1}\exp[-(2/n)\xi_i]
\end{eqnarray}

\noindent where ${\cal{M}}(1)$ is the normalization constant found in the prior section, $H[u]$ is the heaviside function, and $\xi_i$ is the $i$th root that satisfies $\lambda({\xi})=\lambda$. The above is straightforward to interpret. The roots $\xi_i$ label shell crossings at a particular scale and the exponential factor accounts for the mass difference between shells that turn around at different times. 

Since the trajectory and the mass profile depend on each other, it is necessary to first assume a mass profile, then calculate the trajectory from eq.\ (\ref{EOMSS2}) and resulting mass profile from eq.\ (\ref{Mass}) and repeat until convergence is reached. 

The density profile $D(\lambda)$ is straightforward to derive using eqs.\ {\ref{rhoDef}}, {\ref{MDef}} and {\ref{Mass}}. We find \cite{Bert,FG}: 

\begin{eqnarray}\label{Density}
D(\lambda)&=&\frac{1}{3\lambda^2}\frac{d{\cal{M}}}{d\lambda} \nonumber \\
&=& \frac{2}{3n}\frac{{\cal{M}}(1)}{\lambda^2}\sum_i(-1)^i\exp[-(2/n)\xi_i]\left(\frac{d\lambda}{d\xi}\right)^{-1}_i \nonumber \\
\end{eqnarray}

\section{Asymptotic Behavior}
\label{sec:AB}

Unlike N-body experiments, self-similar systems are not limited by resolution. One can analytically infer the asymptotic slope of the mass profile close to the origin. FG did this by taking advantage of adiabatic invariance, self-consistently calculating the mass profile, and analyzing the limit of the mass profile as $\lambda\rightarrow0$. Below, we generalize their analysis to the case of particles with changing angular momentum. Unlike Nusser \cite{Nusser}, we do not restrict our analysis to the case $\varpi=0$. 

Just as in FG, we start by parameterizing the halo mass and the variation of the apocenter distance $r_a$.

\begin{eqnarray}
M(r,t)&=&\kappa(t)r^{\alpha} \label{MassParam}\\
\frac{r_a}{r_*}&=&\left(\frac{t}{t_*}\right)^q \label{ra}
\end{eqnarray}

\noindent In the above $r_*$ is the turnaround radius of a mass shell which turns around at $t_*$. It is possible to relate $q$ and $\alpha$ to $n$ by taking advantage of adiabatic invariance. The equation of motion for the mass shell is:

\begin{equation}
\frac{d^2r}{dt^2}=-G\kappa(t)r^{\alpha-2}+\frac{L^2(t)}{r^3}
\end{equation}

At late times, the orbital period is much smaller than the time scale for the mass and angular momentum to grow. Taking $\kappa(t)$ and $L(t)$ to stay roughly constant over an orbit and integrating the above equation, we find the energy equation:

\begin{equation}
\left(\frac{dr}{dt}\right)^2= \frac{2G\kappa(t)}{\alpha-1}(r_a^{\alpha-1}-r^{\alpha-1})-L^2(t)(r^{-2}-r_a^{-2})
\end{equation} 

The above relationship tells us how the pericenters $r_p$ evolve with time. Note that we only consider torquing models (eq.\ {\ref{torque}}) which give rise to bound orbits. This restriction on $\varpi$ will be discussed below. Defining $y\equiv r_p/r_a$ and evaluating the above at $r=r_p$, we find:

\begin{equation}\label{peri}
\frac{1-y^{\alpha-1}}{y^{-2}-1}\equiv A(y) =\frac{(\alpha-1)L^2(t)}{2G\kappa(t)r_a^{\alpha+1}(t)}
\end{equation}

\noindent For $\varpi > (<)\; 0$, the angular momentum of particles in the mass shell increases (decreases). This gives rise to pericenters that increase (decrease). Hence at late times, the orbit of a mass shell with increasing angular momentum will circularize and have $y\sim1$, while the orbit of a mass shell with decreasing angular momentum will become more radial, with $y\ll1$. With this in mind, we can now calculate the radial action in order to find how $q$ relates to $n$ and $\alpha$. The radial action is given by:

\begin{eqnarray}
J&=&2\int^{r_a}_{r_p}dr\left(\frac{dr}{dt}\right) \nonumber \\
&=& 2\left(\frac{2G\kappa(t)}{\alpha-1}\right)^{1/2}r_a^{(\alpha+1)/2} \times\nonumber \\
&&\int^1_{y(t)}du\left[(1-u^{\alpha-1})-A(y)(u^{-2}-1)\right]^{1/2}
\end{eqnarray}

In the above we've assumed $\alpha>1$. Generalizing to the case $\alpha<1$ is straightforward. The special case $\alpha=1$ will be addressed later. For $y(t)\ll1$, the above integral is dominated by the region in which $y(t)\ll u \ll 1$. Over this region, the integrand is time independent and hence the same for all orbits. Therefore adiabatic invariance implies $\kappa(t)r_a^{\alpha+1} = \rm{const}$. For $y(t)\sim1$, the orbit is circular, which implies the radial action vanishes and $L^2(t)=G\kappa(t)r_a^{\alpha+1}$.  
Using eq.\ (\ref{ra}), and noting that $\kappa(t)\propto t^{s}$ where $s=3\beta-2-\alpha\beta$, we find at late times: 

\begin{equation}\label{q}
q=
\begin{cases}
\frac{1}{\alpha+1}\{2\varpi+\frac{2}{3n}[\alpha(1+n)-3]\} & \text{if $\varpi\geq0$} \\
\frac{2}{3n(\alpha+1)}\left[\alpha(1+n)-3\right]  & \text{if $\varpi<0$}
\end{cases}
\end{equation} 

\noindent For the specific case, $\varpi<0$, taking advantage of $y\ll1$, the adiabatic invariance arguments above, and eqs.\ (\ref{LDef}) and (\ref{torque}), we can rewrite eqn.\ (\ref{peri}) in the form $y(t,t_*)=y_0(t/t_*)^l$, where: 

\begin{equation} \label{peri2}
l= 
\begin{cases} 
\varpi & \text{if $\alpha>1$}, \\ 
2\varpi/(\alpha+1) & \text{if $\alpha<1$}. 
\end{cases} 
\end{equation} 

\noindent and $y_0r_*$ is the pericenter of a mass shell at turnaround. Constant angular momentum after turnaround corresponds to $\varpi=0$. This case was addressed analytically in \cite{Nusser} and numerically in \cite{Sikivie}. 

We next take advantage of the functional form of the mass profile. Following FG, we define $P(r/r_a,y)$ to be the fraction of time a particle with apocenter distance $r_a$ and pericenter $yr_a$, at a particular time $t$, spends inside $r$.

\begin{eqnarray}
P(v,y)&=&0\quad\quad\;\; (v < y) \nonumber \\
P(v,y)&=&\frac{I(v,y)}{I(1,y)} \quad (y< v\leq1) \nonumber \\
P(v,y)&=&1\quad\quad\;\; (v>1)
\end{eqnarray}

\noindent where

\begin{equation} \label{Ivy}
I(v,y)\equiv
\begin{cases}
\int^v_y\frac{du}{\left((1-u^{\alpha-1})-A(y)(u^{-2}-1)\right)^{1/2}} & \text{if $\alpha>1$}, \\ 
\int^v_y\frac{du}{\left((u^{\alpha-1}-1)+A(y)(u^{-2}-1)\right)^{1/2}} & \text{if $\alpha<1$}.
\end{cases}
\end{equation}

\noindent We see that the presence of pericenters causes the new case $v<y$, which did not exist in the FG analysis. Self consistency demands that

\begin{equation}
\left(\frac{r}{r_{\rm{ta}}}\right)^{\alpha}=\frac{M(r,t)}{M(r_{\rm{ta}},t)}=\int^{M_{ta}}_0\frac{dM_*}{M_{ta}}P\left(\frac{r}{r_a(t,t_*)},y(t,t_*)\right) \\
\end{equation}

\noindent where $M_*$ is the mass internal to a shell that turns around at $t_*$ and $M_{ta}$ is the current turnaround mass. The integral assigns a weight to each shell depending on how often that shell is below the scale $r$. Noting from eq.\ (\ref{MDef}) that 

\begin{equation}
M_*=M_{ta}\left(\frac{t_*}{t}\right)^{3\beta-2},
\end{equation}

\noindent using eq.\ (\ref{ra}) and transforming integration variables, we find:

\begin{equation}\label{MassCons}
\left(\frac{r}{r_{\rm{ta}}}\right)^{\alpha-k}=k\int^{\infty}_{r/r_{\rm{ta}}}\frac{du}{u^{1+k}}P\left(u,y(t,t_*)\right)
\end{equation}

\noindent where 

\begin{equation}\label{k}
k=\frac{6}{2+n(2-3q)}
\end{equation}

\noindent As $u$ increases, the above integral sums over shells with smaller $t_*$. Since the pericenter of a shell evolves with time, the second argument of $P$ depends on $u$. The dependence, as we showed, varies with torque model (sign of $\varpi$); hence we've kept the dependence on $u$ implicit. Next we analyze the above for certain regimes of $r/r_{\rm{ta}}$, and certain torquing models, in order to constrain the relationship between $\alpha$ and $k$. 

\subsection{$r/r_{\rm{ta}} \ll y_0$, $\varpi<0$}

For $\varpi<0$, particles lose angular momentum over time. When probing scales $r/r_{\rm{ta}}\ll y_0$, mass shells with $t_*\ll t_{ta}$ only contribute. As a result, $y(t,t_*)\ll1$. Using eq.\ (\ref{peri2}), we then find:

\begin{equation}\label{peri3}
y(t,t_*)=y_0\left(\frac{t}{t_*}\right)^l=y_0\left(\frac{r}{ur_{\rm{ta}}}\right)^\delta
\end{equation}

\noindent where $\delta\equiv l/(q-\beta)$. For bound mass shells, $q-\beta<0$. Therefore, since $\delta>0$, the first argument of $P$ in eq.\ (\ref{MassCons}) increases while the second decreases as we sum over shells that have turned around at earlier and earlier times ($u\rightarrow\infty$). For $r/r_{\rm{ta}} \ll y_0$, mass shells which most recently turned around do not contribute to the mass inside $r/r_{\rm{ta}}$ since we are probing scales below their pericenters. Mass shells only begin to contribute when the two argument of $P$ are roughly equal to each other. This occurs around:

\begin{equation}\label{yprime}
u = y_1\equiv \left(y_0(r/r_{\rm{ta}})^{\delta}\right)^{1/(1+\delta)}
\end{equation}

\noindent Hence, we can replace the lower limit of integration in eq.\ (\ref{MassCons}) with $y_1$. We next want to calculate the behavior of eq.\ (\ref{MassCons}) close to $y_1$ in order to determine whether the integrand is dominated by mass shells around $y_1$ or mass shells that have turned around at much earlier times. The first step is to calculate the behavior of $P(u,y)$ for $u\approx y$. We find:

\begin{equation}
P(u,y)\propto u^{1/2}(1-y/u)^{1/2} \times 
\begin{cases} y^{1/2} & \text{if $\alpha>1$}, \\ 
y^{1-\alpha/2} & \text{if $\alpha<1$}.
\end{cases}
\end{equation}
   
\noindent Given the above, we evaluate the indefinite integral in eq.\ (\ref{MassCons}), noting that $y$ is a function of $u$ (eq.\ {\ref{peri3}}). For $u\sim y_1$, we find:

\begin{eqnarray}\label{integral}
\int\frac{du}{u^{1+k}}&P&\left(u,y_0\left(\frac{r}{ur_{\rm{ta}}}\right)^{\delta}\right)\nonumber \\
&\propto& (u/y_1-1)^{3/2}
\begin{cases}
y_1^{1-k} & \text{if $\alpha>1$}, \\
y_1^{3/2-k-\alpha/2} & \text{if $\alpha<1$}. \nonumber 
\end{cases} \\
\end{eqnarray} 

Now comes the heart of the argument. Following the logic in FG, if we keep $u/y_1$ fixed and the integrand blows up as $r/r_{\rm{ta}} \rightarrow 0$, then the left hand side of eq.\ (\ref{MassCons}) must diverge in the same way as the right hand side shown in eq.\ (\ref{integral}). Therefore, using eq.\ (\ref{yprime}):

\begin{equation}\label{diverge}
\alpha-k= 
\begin{cases}
\delta(1-k)/(1+\delta)& \text{if $\alpha>1$}, \\ 
\delta(3/2-k-\alpha/2)(1+\delta) & \text{if $\alpha<1$}.
\end{cases}
\end{equation} 

\noindent Otherwise, if the right hand side converges, then the integrand will not depend on $r/r_{\rm{ta}}$ as $r/r_{\rm{ta}}\rightarrow 0$. Therefore, the left hand side cannot depend on $r/r_{\rm{ta}}$ either, which implies $\alpha=k$. Solving the above system of equations for $\alpha$ given eqs.\ (\ref{q}) and (\ref{k}) and making sure the solution is consistent, (ie: using eq.\ (\ref{diverge}) only if the integrand diverges), we find:

\begin{eqnarray}\label{relations1}
\text{For }n&\leq&2: \nonumber \\
\alpha&=&\frac{1+n-\sqrt{(1+n)^2+9n\varpi(n\varpi-2)}}{3n\varpi} \nonumber \\
k&=&\frac{1+n+3n\varpi-\sqrt{(1+n)^2+9n\varpi(n\varpi-2)}}{n\varpi(4+n)} \nonumber \\
q&=&\frac{1+n-3n\varpi-\sqrt{(1+n)^2+9n\varpi(n\varpi-2)}}{3n} \nonumber \\ 
\text{For }n&\geq&2: \nonumber \\
\alpha&=&k=\frac{3}{1+n}\;, \quad q=0
\end{eqnarray}

\noindent The above solutions are continuous at $n=2$. Moreover, taking the no-torque limit ($\varpi\rightarrow0$) for $n\leq2$ gives the same solutions as $n\geq2$, which is consistent with analytic and numeric results from \cite{Nusser,Sikivie}. Taking the limit, $\varpi\rightarrow-\infty$ reproduces the FG solution, as expected, since the shell loses its angular momentum instantly. The solution for $n\geq2$ is independent of $\varpi$. This is because the mass is dominated by shells with turnaround time $t_*\ll t_{ta}$ which have effectively no angular momentum. In other words, for $\varpi<0$, the solution should only depend on torquing parameters when the mass is dominated by shells that have turned around recently. 
 
\subsection{$r/r_{\rm{ta}} \ll y_0$, $\varpi>0$}

For $\varpi>0$, the angular momentum of particles increase with time. As mentioned above, when probing scales $r/r_{\rm{ta}} \ll y_0$, mass shells with $t_* \ll t_{ta}$ only contribute. As a result, $y(t,t_*)\sim1$. In other words, the orbits are roughly circular. We can therefore replace the lower limit of integration in eq.\ (\ref{MassCons}) with 1 since mass shells will only start contributing to the sum when $u\sim y\sim1$. Hence, the right hand side of eq.\ (\ref{MassCons}) does not depend on $r/r_{\rm{ta}}$, which implies $\alpha=k$. Using eq.\ (\ref{q}) and (\ref{k}), we find:

\begin{equation}\label{relations3}
\alpha=k=\frac{3}{1+n-3n\varpi}\;, \quad q=2\varpi\;, \quad\quad\; \text{for}\; 0\leq n\leq3
\end{equation} 

The no torque case, $\varpi=0$, is consistent with the analysis in the prior subsection. The singularity $\varpi = (1+n)/3n$ implies $q=\beta$. This physically corresponds to the orbital radius of mass shells increasing at the same rate as the turnaround radius and results in orbits that are not bound and a cored profile where there are no particles internal to a particular radius. This breaks the assumption of a power law mass profile (eq.\ \ref{MassParam}); hence we only consider $\varpi<(1+n)/3n$.

Unlike the Nusser solution, certain parameters give $\alpha>3$, which corresponds to a density profile which converges as $r\rightarrow 0$. Since the angular momentum acts like a heat source $d\rho/dr>0$ is dynamically stable and physical.  
 
\subsection{$y_0\ll r/r_{\rm{ta}} \ll1$}

In this regime, we are probing scales larger than the pericenters of the most recently turned around mass shells. As a result, $P(u,y)$ is dominated by the contribution from the integrand when $u\gg y$. Therefore: 

\begin{equation}
P(u,y)\propto 
\begin{cases} u & \text{if $\alpha>1$}, \\ 
u^{(3-\alpha)/2} & \text{if $\alpha<1$}.
\end{cases}
\end{equation}

\noindent Hence the integral in eq.\ (\ref{MassCons}) becomes:

\begin{eqnarray}\label{integral2}
\int\frac{du}{u^{1+k}}P\left(u,y(t,t_*)\right)\nonumber 
\propto
\begin{cases}
u^{1-k} & \text{if $\alpha>1$}, \\
u^{3/2-k-\alpha/2} & \text{if $\alpha<1$}. \nonumber 
\end{cases} \\
\end{eqnarray} 

\noindent Following the logic in the prior section, if the integral diverges as $r/r_{\rm{ta}}\rightarrow0$, then we set the exponents on the left hand side and right hand side equal to each other so that both sides diverge in the same way. If the integral converges, then the left hand side cannot depend on $r/r_{\rm{ta}}$, which implies $\alpha=k$. Given these arguments, and imposing consistency with the above inequalities on $\alpha$ to find the appropriate ranges for $n$, we find:

\begin{eqnarray}\label{relations2}
\alpha&=&1\;, \quad k=\frac{6}{4+n}\;, \quad q=\frac{n-2}{3n}\;, \quad \text{for}\; n\leq2 \nonumber \\
\alpha&=&k=\frac{3}{1+n}\;, \quad q=0\;, \quad\quad\quad\quad\quad\; \text{for}\; n\geq2 \nonumber \\
\end{eqnarray}

The above is exactly the FG solution. We expect to recover these solutions since we are probing scales larger than the pericenters of the most massive shells, where the angular momentum does not affect the dynamics.  

This section assumed $\alpha\neq1$ and yet, for certain parts of parameters space,  eqs.\ (\ref{relations1}), (\ref{relations3}) and (\ref{relations2}) give $\alpha=1$. However, since the solutions are continuous as $\alpha\rightarrow1$ from the left and right, then the results hold for $\alpha=1$ as well. 

\section{Structure of the Halo}
\label{sec:structure}

In this section, we discuss the radial structure of galactic size halos and compare directly to numerical N-body simulations. Note however, that the mass of a halo is not well defined when our model is applied to cosmological structure formation since it is unclear how the spherical top hat mass which characterizes the halo when it is linear relates to the virial mass which characterizes the halo when it is nonlinear. For halos today with galactic size virial masses, we assume the model parameter $n$ which characterizes the initial density field, is set by a spherical top hat mass of $10^{12}$$M_{\odot}$. As described in prior sections, specifying the top hat mass also sets model parameters $B$ and $p$. Before comparing directly to N-body simulations, we first describe how $\varpi$ influences the halo. 

\begin{figure}
  \begin{center}
    \includegraphics[height = 70mm, width=\columnwidth]{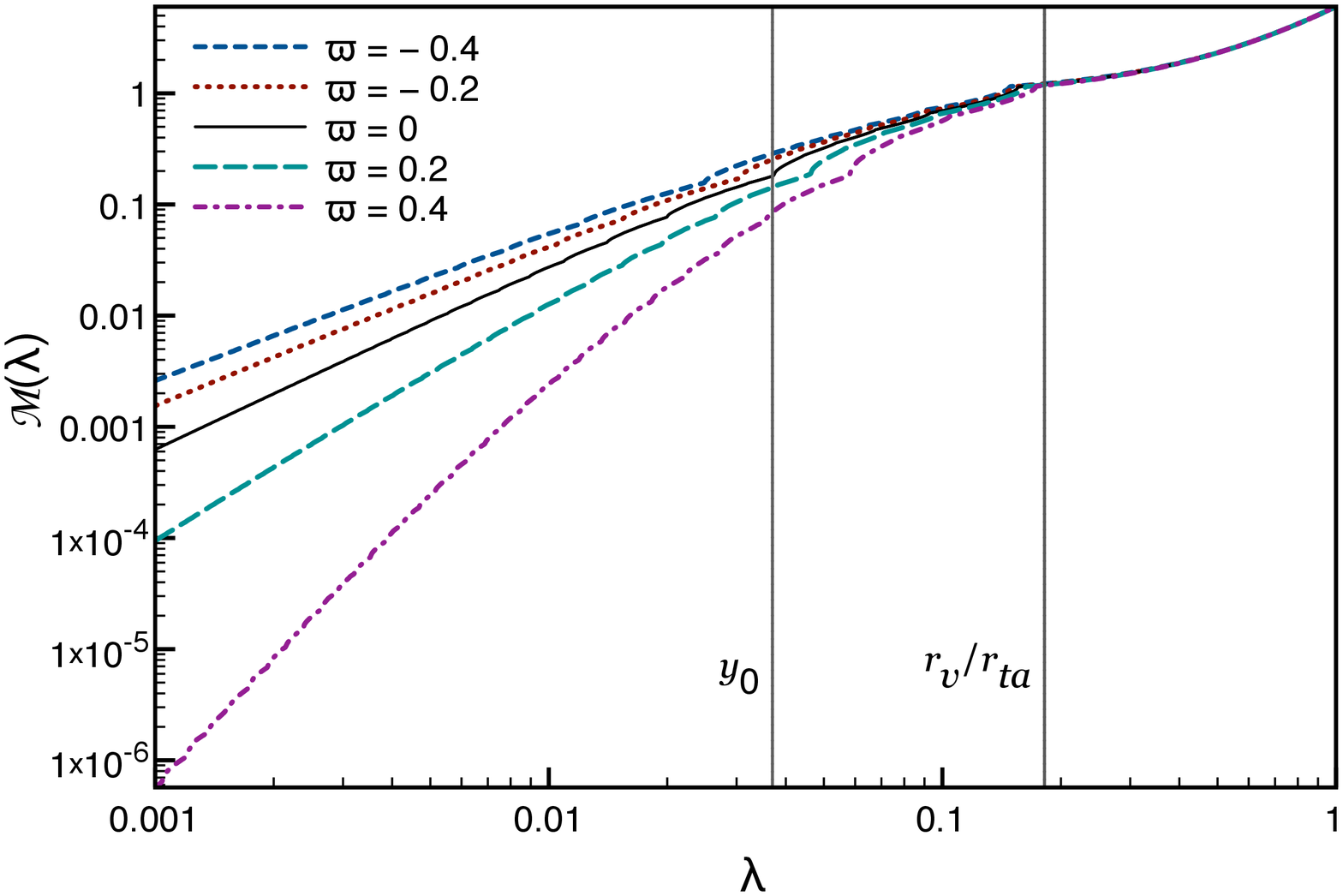}
     \includegraphics[height = 70mm, width=\columnwidth]{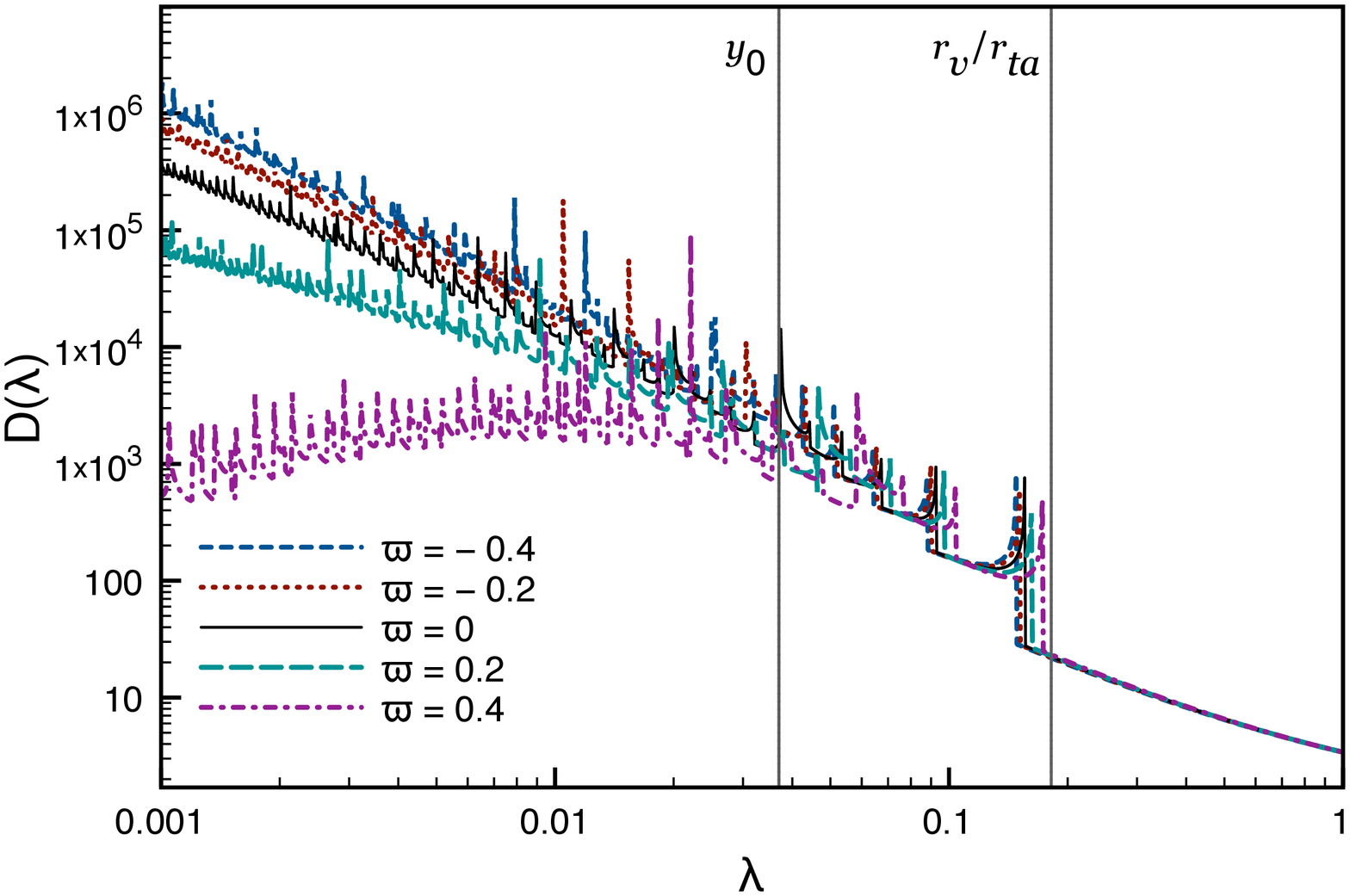}
  \end{center}
\caption{The mass and density profiles for galactic size halos $\{n=0.77,p=2n,B_{1.5}=0.39\}$ with varying $\varpi$. The value of $\varpi$ changes how pericenters evolve with time and thereby affects how many shells at a particular scale contribute to the internal mass. The above numerically computed profiles match analytic predictions. The virial radius ($r_v$) and first pericenter passage ($y_0r_{ta}$) are labeled for clarity. \label{fig:M1}}
\end{figure}

Figure \ref{fig:M1} shows the mass ${\cal{M}}(\lambda)$ and density profiles $D(\lambda)$ for galactic size halos $\{n=0.77,p=2n,B_{1.5}=0.39\}$ with varying $\varpi$. The spikes in the density profile are caustics which form at the shell's turning points. They form because of unphysical initial conditions; we assume each shell has zero radial velocity dispersion. The structure of the halos naturally break down into three different regions. The dividing points between these regions are roughly the virial radius $(r_v)$ and $y_0r_{\rm{ta}}$, the pericenter of the mass shell which most recently turned around.

As in dark matter N-body simulations, we associate the virial radius with $r_{200}$, the radius at which $M(r_{200})=800\pi \rho_B r^3_{200}/3$ is satisfied. Numerically, we find that the virial radius occurs near the first caustic ($\lambda\sim .18$). For $r>r_v$, the mass profile flattens and then starts to increase. The flattening is equivalent to what is seen on large scales in N-body simulations where $\rho\propto r^{-3}$. The mass profile then starts to increase again because at large radii, where $\lambda\gg1$, the density is roughly constant, which implies ${\cal{M}}\propto\lambda^{3}$. For $r\sim r_v$, it is difficult to make analytic predictions for the mass profile because adiabatic invariance breaks down. In other words, the mass of the halo and angular momentum of a shell change on the same time scale as the shell's orbital period.

As discussed in the prior section, for $y_0r_{\rm{ta}}\ll r\ll r_v$, we can take advantage of adiabatic invariance to infer the logarithmic slope of the mass profile. Since this regime probes a scale much larger than the pericenters of the mass shells, the angular momentum does not affect the dynamics and we recover the FG solution. For our particular choice of $n=0.77$, this gives an isothermal profile with $\rho\propto r^{-2}$. However, since $n<2$ for all collapsed objects today (Figure \ref{fig:n}) and based on the results of FG, our model predict that all halos are isothermal in this regime. 

Last, for $r/r_{\rm{ta}}\ll y_0$, angular momentum begins to play a role and the halo starts to exhibit different features than the FG solution. The behavior is very intuitive. The mass of a particular shell does not contribute to the internal mass when probing radii less than the pericenter of that mass shell. Therefore, as one probes radii smaller than the pericenter of the most recently turned around mass shell, one expects a steeper fall off than the FG solution, since less mass is enclosed interior to that radius. Moreover, varying $\varpi$ varies the pericenter of mass shells over time. Increasing (decreasing) angular momentum, $\varpi>(<)\;0$, causes the pericenters to increase (decrease) over time. This results in profiles which are steeper (shallower) than the no-torque case.    

\begin{figure}
  \begin{center}
    \includegraphics[height = 70mm, width=\columnwidth]{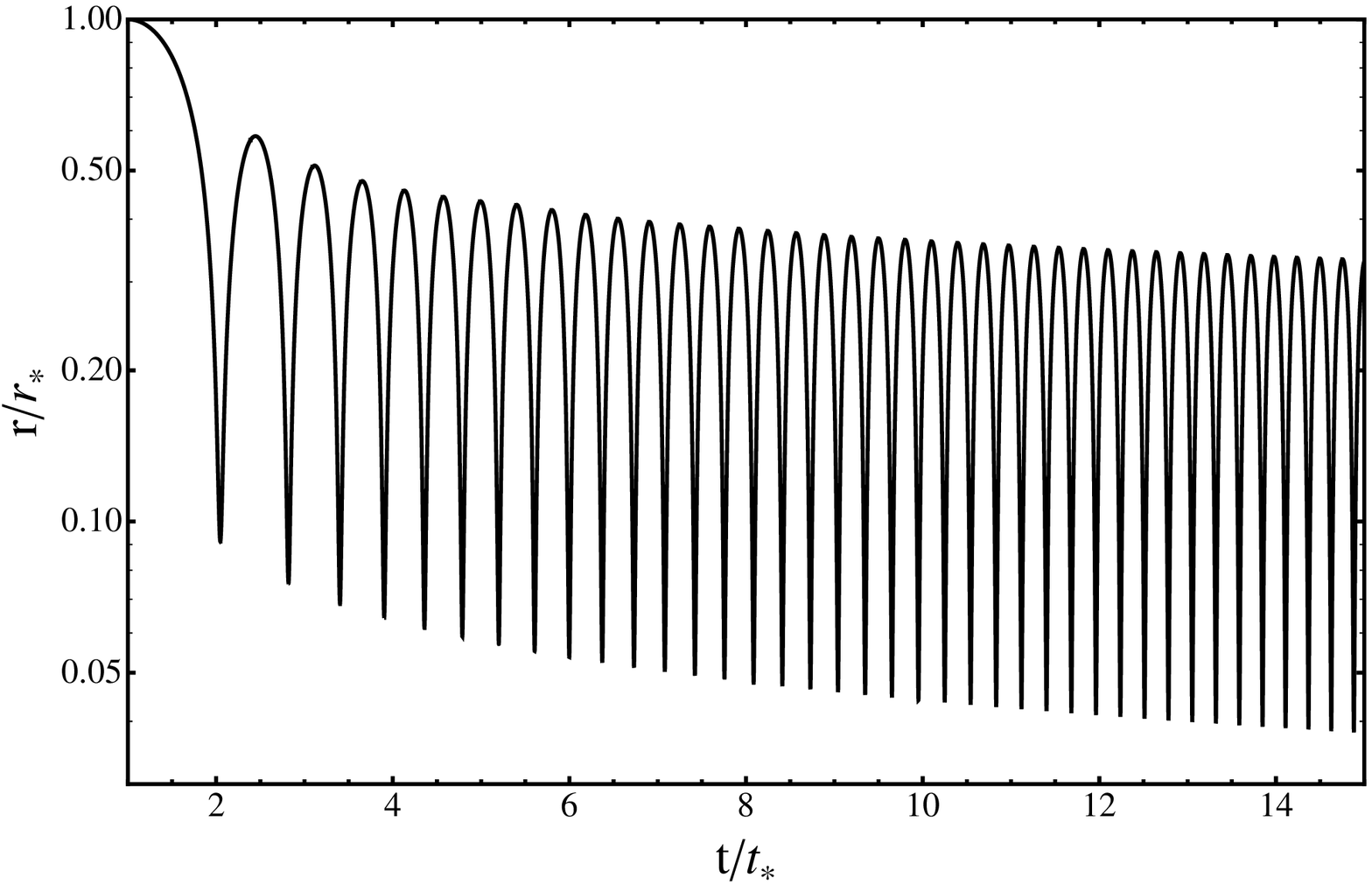}
     \includegraphics[height = 70mm, width=\columnwidth]{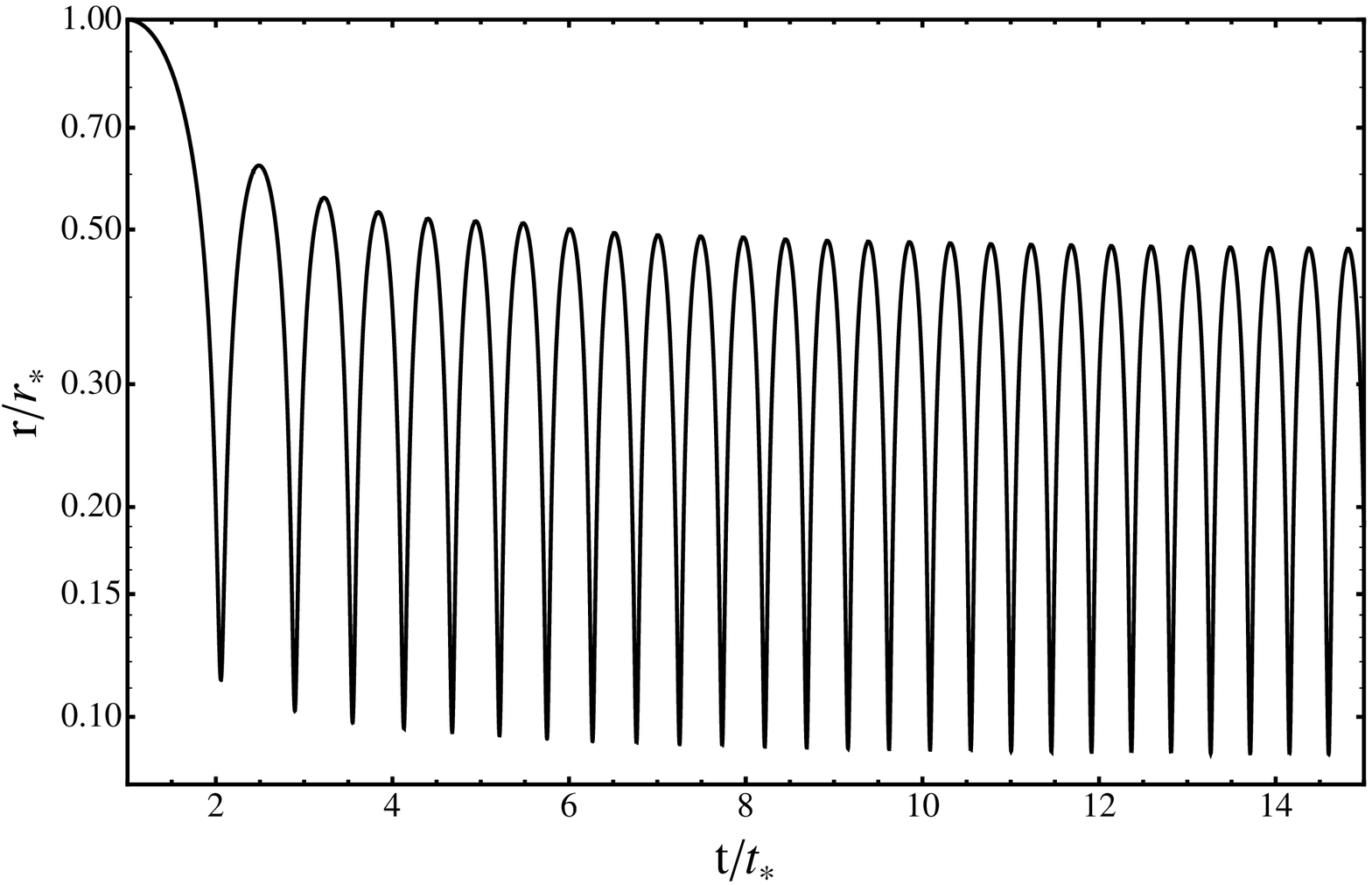}
    \includegraphics[height = 70mm, width=\columnwidth]{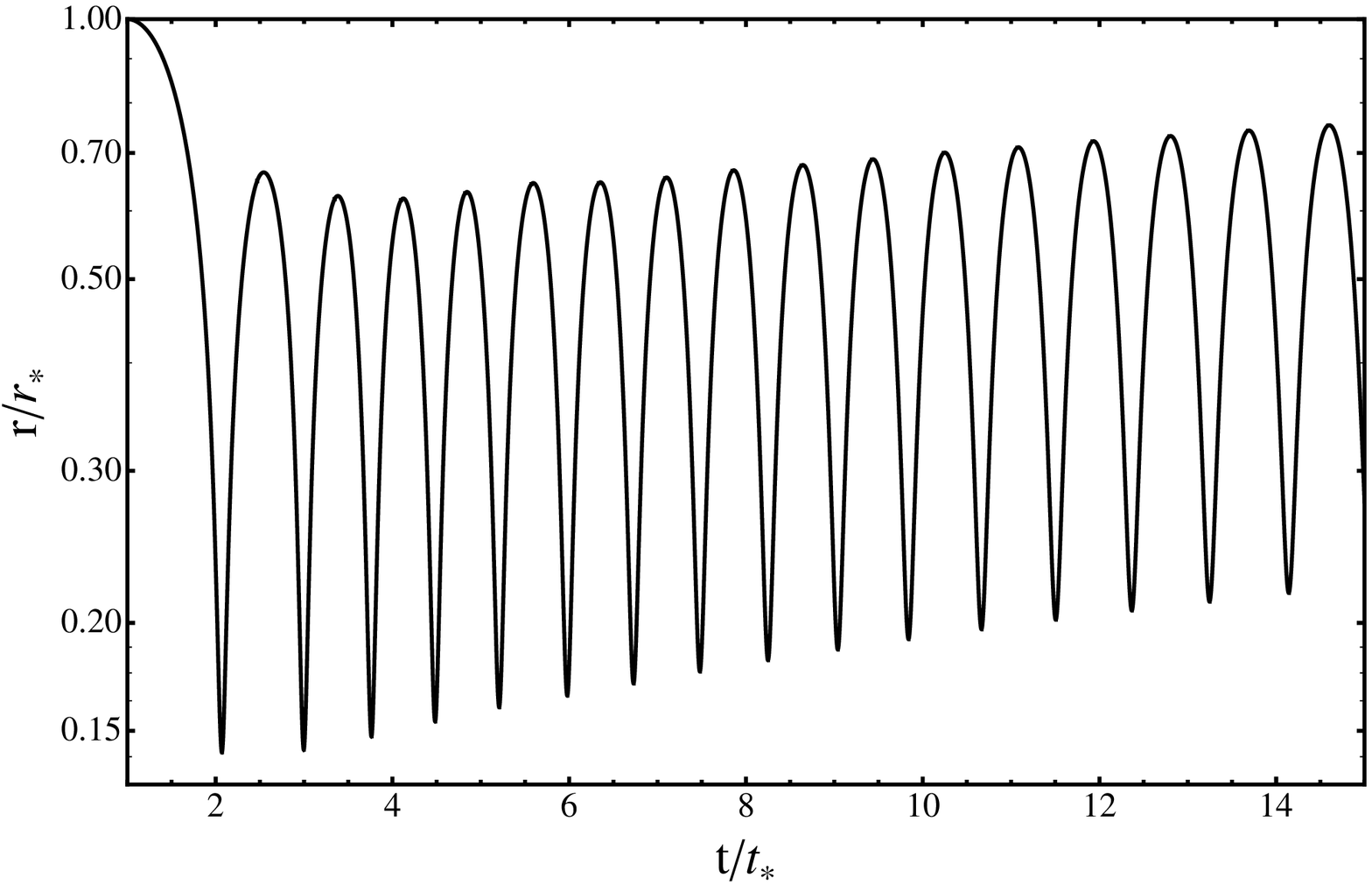}
  \end{center}
\caption{The radius of a mass shell, normalized to its turnaround radius, for a galactic size halo $\{n=0.77,p=2n,B_{1.5}=0.39\}$, plotted vs time. The top panel shows a shell with particles that have decreasing angular momentum ($\varpi=-0.2$). The middle panel shown a shell with particles that have constant angular momentum ($\varpi=0$). The bottom panel shows a shell with particles that have increasing angular momentum ($\varpi=0.2$).} \label{fig:L1}
\end{figure}

\begin{figure}
  \begin{center}
    \includegraphics[height = 70mm, width=\columnwidth]{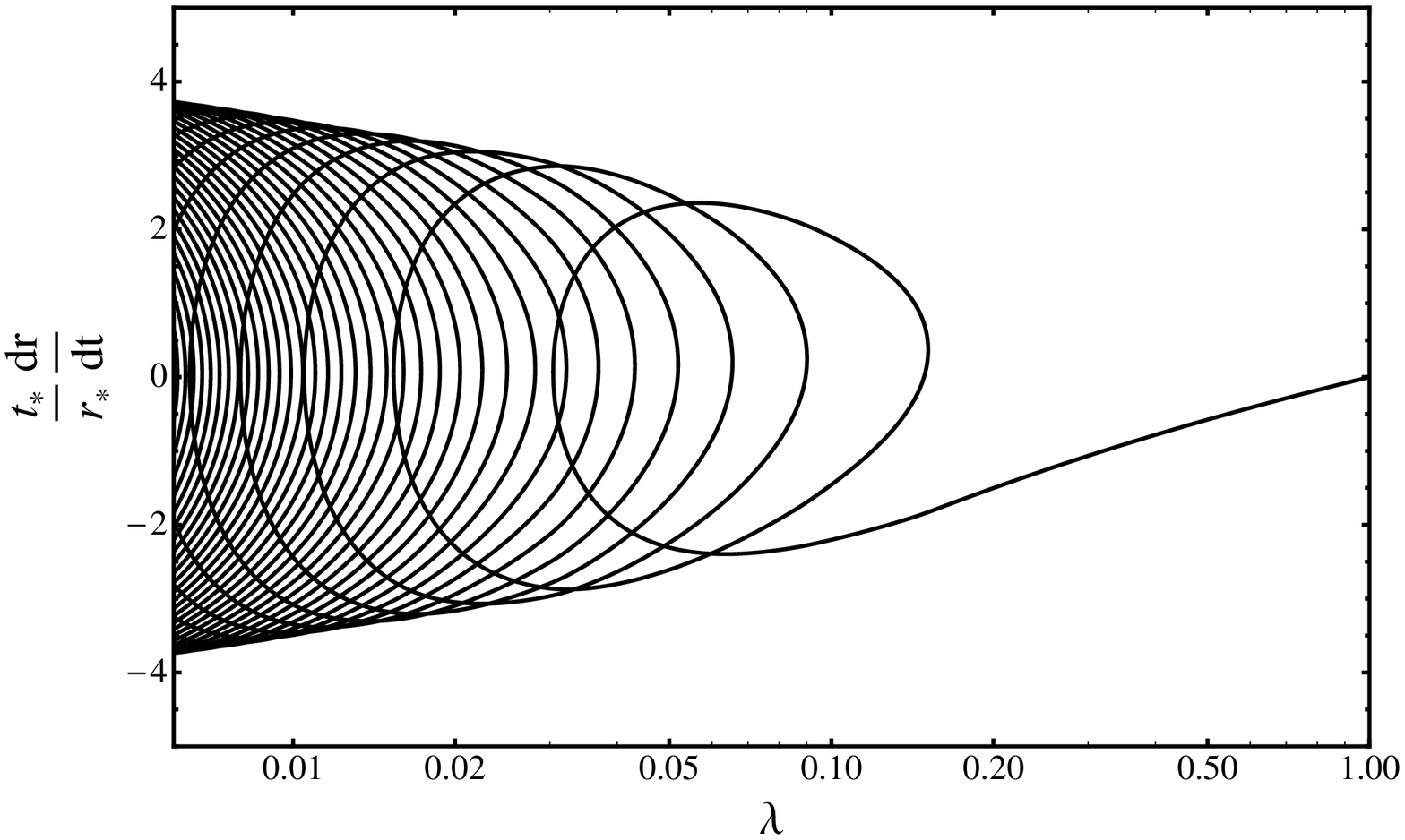}
    \includegraphics[height = 70mm, width=\columnwidth]{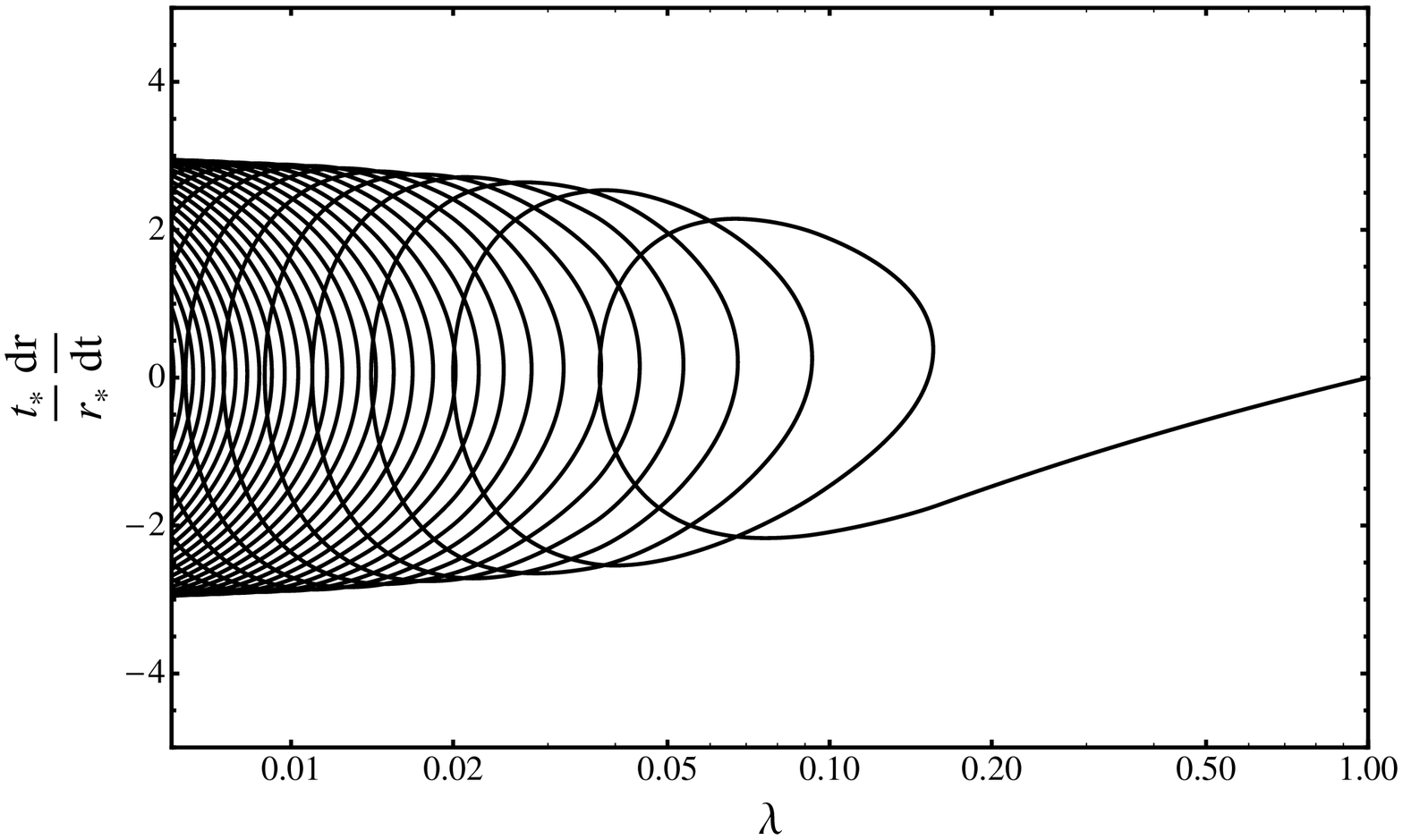}
    \includegraphics[height = 70mm, width=\columnwidth]{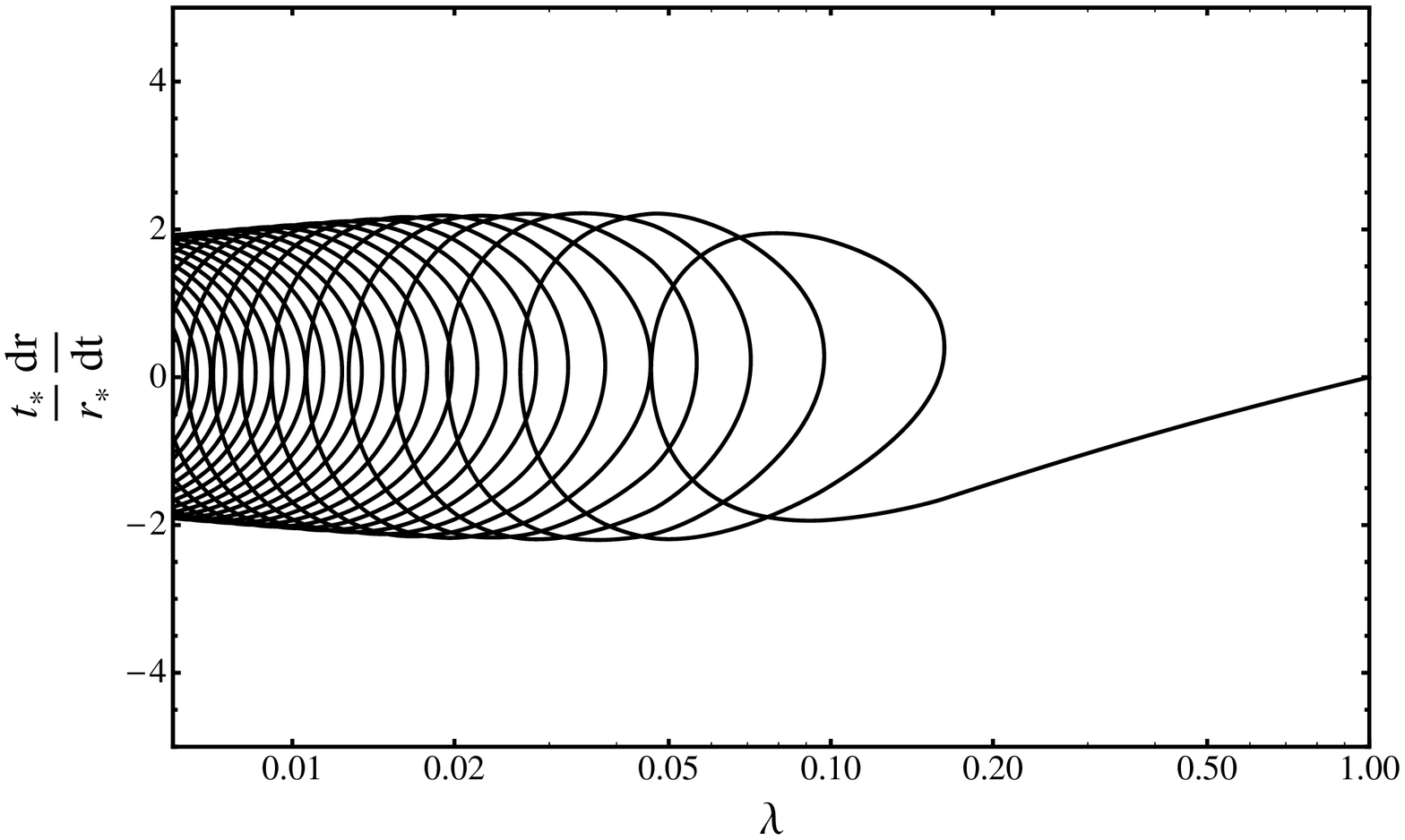}
  \end{center}
\caption{A phase space diagram of a galactic size halo $\{n=0.77,p=2n,B_{1.5}=0.39\}$ at the current turnaround time. Velocities are normalized to the turnaround time and radius of each shell. The top panel shows a halo with $\varpi=-0.2$. The middle panel shows a halo with $\varpi=0$. The bottom panel shows a halo with $\varpi=0.2$.} \label{fig:PH1}
\end{figure}

It is informative to find how the transition radius $y_0$ depends on model parameters. Since the mass and angular momentum grow significantly before the first pericenter, we can only approximately determine this relationship. We assume that the profile is isothermal on large scales and the halo mass and shell angular momentum are fixed to their turnaround values. For $y_0\ll1$, the transitional radius $y_0$ solves the transcendental equation $y_0^2\ln(y_0)=-9B^2/4{\cal{M}}(1)$. As expected, $B\rightarrow0$ reproduces $y_0\rightarrow0$. As shown in Figure \ref{fig:contour}, ${\cal{M}}(1)$ varies with $\{B,p\}$. However, for reasonable parameter values, the mass normalization is changed at most by a factor of 2. Therefore, $y_0$ most strongly depends on $B$, and $p$ has a negligible effect on the structure of the halo. As seen in figure \ref{fig:M1}, $y_0$ should also depends on $\varpi$. The above approximation neglects this dependence since we assumed the angular momentum is set to the turnaround value.  

Figure \ref{fig:L1} shows the radius of a mass shell, normalized to its turnaround radius $r_*$, for a galactic size halo $\{n=0.77,p=2n,B_{1.5}=0.39\}$, as a function of time. In the top panel, particles in the shell lose angular momentum ($\varpi=-0.2$), in the middle panel the angular momentum remains constant ($\varpi=0$), while in the bottom panel particles in the shell gains angular momentum ($\varpi=0.2$). As expected, the pericenters in the top panel decrease with time while the pericenters grow in the bottom panel. Hence, the orbits of particles with decreasing angular momentum become more radial while those with increasing angular momentum become more circular. 

Notice that the period of oscillation also varies for different $\varpi$. The period of oscillation is set by the shell's apocenter $r_a$ and the mass internal to $r_a$. Using the adiabatic invariance relations we found in Section \ref{sec:AB} and assuming Kepler's third law, we find that the period of the orbit $P\propto r^2_a$ for shells with decreasing angular momentum and $P\propto r_a$ for shells with increasing angular momentum. Moreover, from eqs.\ (\ref{relations1}) and (\ref{relations2}), $r_a$ decreases (increases) with time for $\varpi < (>)\;0$. Though Kepler's third law doesn't hold for this system, it still gives intuition for the above results. 

Figure \ref{fig:PH1} shows the phase space diagram for a galactic size halo $\{n=0.77,p=2n,B_{1.5}=0.39\}$. In the top panel, the particles in the shell lose angular momentum ($\varpi=-0.2$), in the middle panel the angular momentum remains constant ($\varpi=0$), while in the bottom panel the particles in the shell gain angular momentum ($\varpi=0.2$). The diagram labels the phase space point of every shell at the current turnaround time. All radial velocities are normalized to the shell's turnaround time $t_*$ and radius $r_*$. Unlike FG, the presence of angular momentum results in caustics associated with pericenters as well, which can be seen in the lower panel of Figure \ref{fig:M1}. In addition, since an increasing angular momentum results in increasing pericenters, the pericenter caustics are more closely spaced in the lower panel than in the upper panel. Moreover, the amplitude of the radial velocity is smaller in the lower panel because orbits are circularizing. The phase space curve appears to intersect itself because we did not plot the tangential velocity component. In full generality, the distribution in the phase space $(r,v_r,v_t=L/r)$ for our model is a non-self-intersecting one-dimensional curve. 

\subsection{Comparing with N-body Simulations}
\label{nbody}
 
\begin{figure}
  \begin{center}
    \includegraphics[height = 70mm, width=\columnwidth]{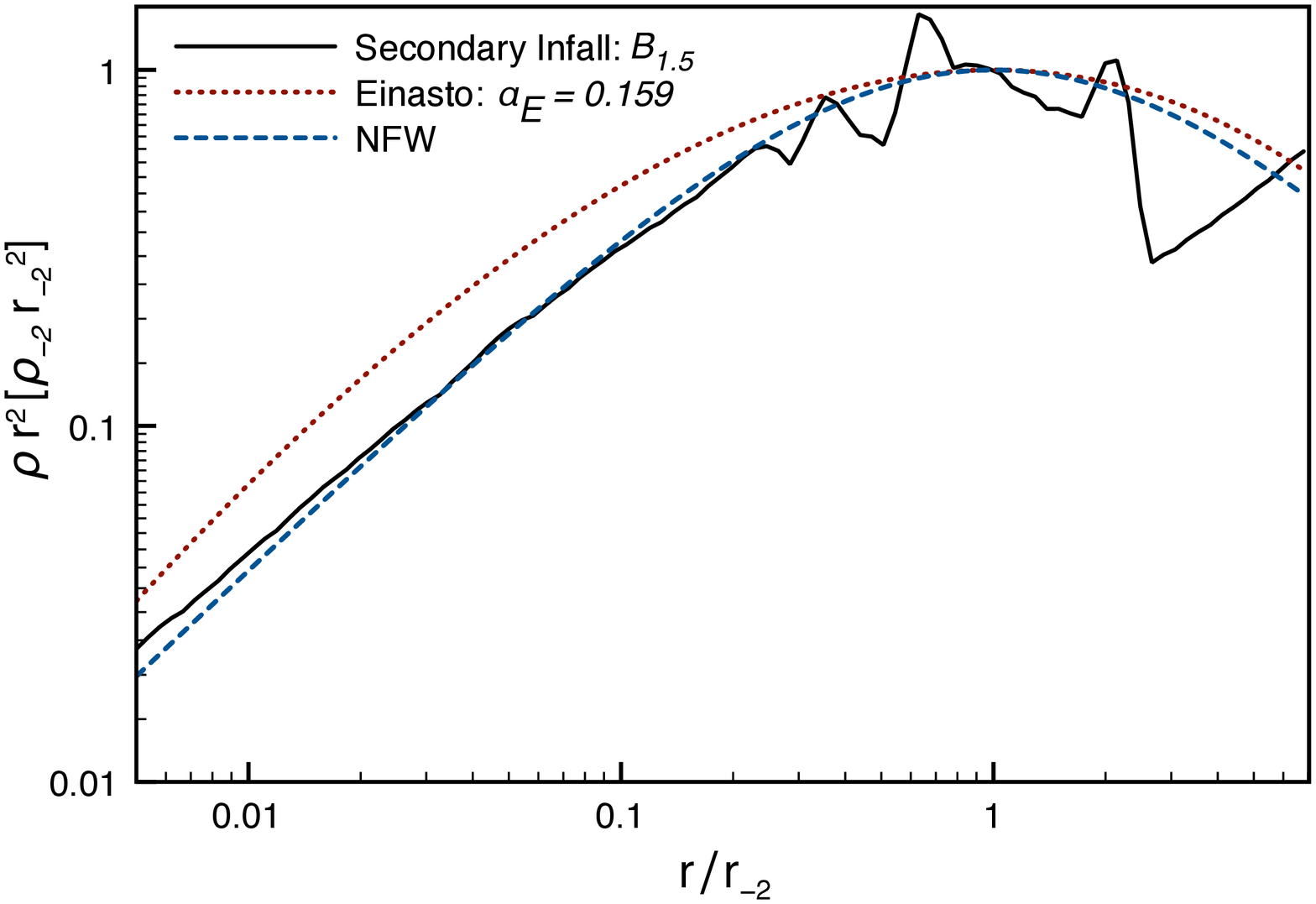}
    \includegraphics[height = 70mm, width=\columnwidth]{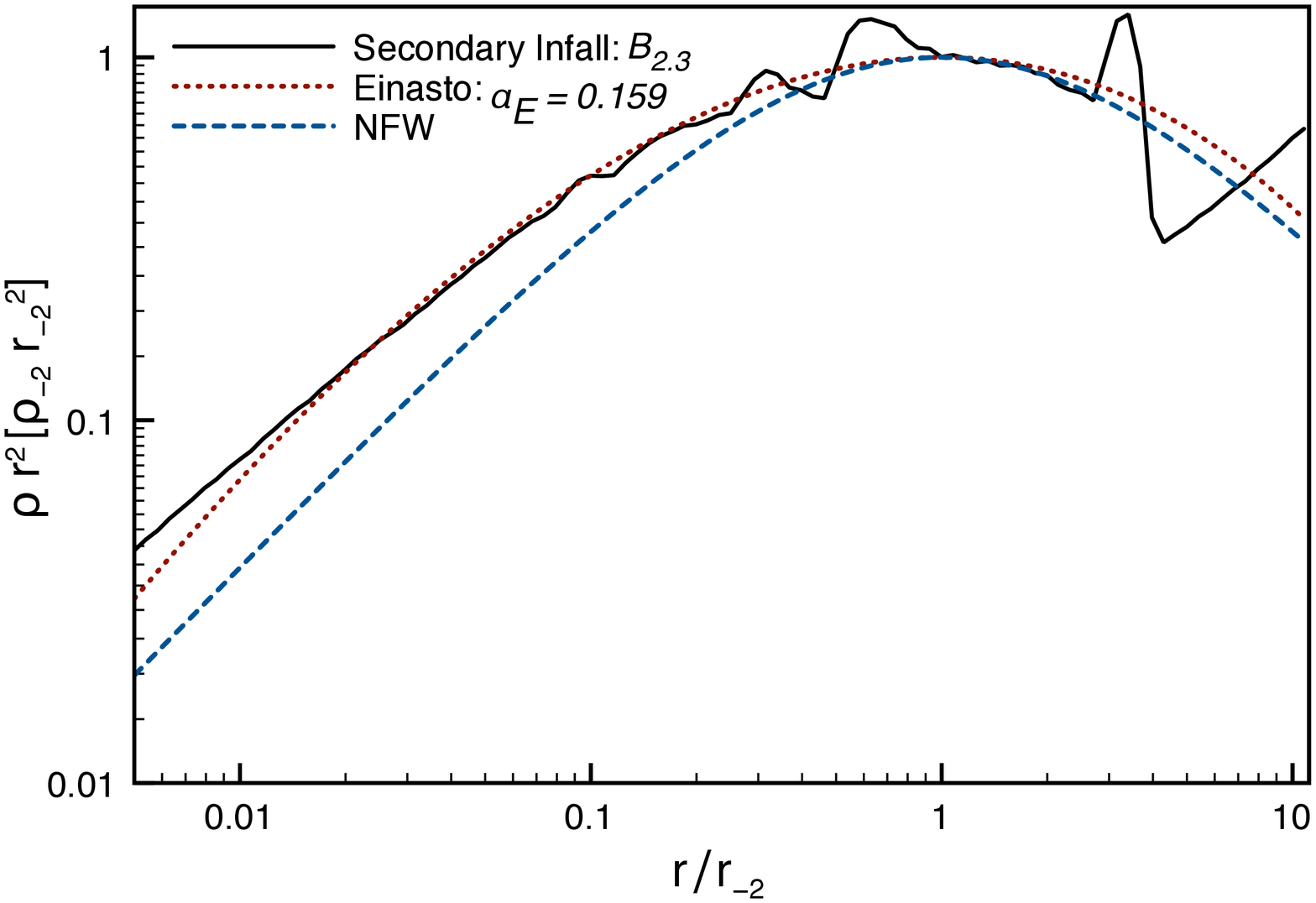}
  \end{center}
\caption{Spherically averaged density profile for the secondary infall model compared with NFW and Einasto profiles. The secondary infall model is calculated for a galactic size halo $\{n=0.77,p=2n\}$ with $\varpi=.1167$. In the top panel we chose $B_{1.5}=0.39$ while in the bottom panel we chose $B_{2.3}=0.26$.} \label{fig:Compare}
\end{figure} 

In this subsection, we compare the density profile of our model's halo to empirical fits inspired by N-body simulations. We first numerically calculate the density profile for a galactic size halo with $\varpi= 0.12$. This value of $\varpi$ was chosen so that $\rho\propto r^{-1}$ on small scales. We then compute the spherically averaged density in 50 spherical shells equally spaced in $\log_{10}r$ over the range $1.5\times 10^{-4} < r/r_v < 3$, and take $r_v=r_{200}$ (defined above). This is the same procedure followed with the recent Aquarius simulation \cite{Aquarius}. Next, we calculate $r_{-2}$, the radius where $r^2\rho$ reaches a maximum. For our halo, as discussed above, the profile is isothermal over a range of $r$. Moreover, the maximum peaks associated with the caustics are unphysical. So, we choose a value of $r_{-2}$ in the isothermal regime that gives good agreement with the empirical fits. Changing $r_{-2}$ does not change our interpretation of the results. 

In Figure \ref{fig:Compare} we compare our spherically averaged density profile to NFW and Einasto profiles. We plot $r^2\rho$ in order to highlight differences. The NFW profile is given by \cite{NFW}: 

\begin{equation}
\rho(r)=\frac{4\rho_{-2}}{(r/r_{-2})(1+r/r_{-2})^2}
\end{equation}

\noindent while the Einasto profile is given by:

\begin{equation}
\ln \Big[\rho(r)/\rho_{-2}\Big]=(-2/\alpha_E)[(r/r_{-2})^{\alpha_E}-1]
\end{equation}

\noindent where $\rho_{-2}$ is the density of our halo at $r_{-2}$ and $\alpha_E$, known as the shape parameter, sets the width of the $r^2\rho$ peak. In the top panel, we use $B_{1.5}=.39$ while in the bottom panel, we use $B_{2.3}=.26$. We choose $\alpha_E=0.159$ since this value was used in Figure 3 of Navarro et. al. \cite{Aquarius}. 

We see that the secondary infall model works surprisingly well. The peaks are a result of the caustics that arise because of cold radial initial conditions. The first spike on the right comes from the first apocenter passage while the second comes from the first pericenter passage. The location of pericenter is most strongly influenced by the model parameter $B$. Hence, the isothermal region is smaller in the top panel than in the bottom panel since particles have less angular momentum at turnaround in the lower panel than in the upper panel. The parameter $B$ then plays the same role as $\alpha_E$; it sets the width of the isothermal region. If we assume N-body simulations faithfully represent dark matter halos, then Figure \ref{fig:Compare} implies that our estimate of $B$ in eq.\ (\ref{pBconstraints}) overestimates the actual value by 1.5 to 2.3. We discuss possible reasons for this in Appendix \ref{sec:TT}. 
  
\section{Discussion} 
\label{sec:disc}

N-body simulations reveal a wealth of information about dark matter halos. Older simulations predict density profiles that are well approximated by an NFW profile \cite{NFW}, while more recent simulations find density profiles that fit better with a modified NFW profile \cite{Moore} or the Einasto profile \cite{Aquarius}. In an attempt to gain intuition for these empirical profiles, we've generalized the self-similar secondary infall model to include torque. This model doesn't suffer from resolution limits and is much less computationally expensive than a full N-body simulation. Moreover, it is analytically tractable. Using this model, we were able to analytically calculate the density profile for $r/r_{\rm{ta}} \ll y_0$ and $y_0\ll r/r_{\rm{ta}}\ll1$. Note that the self-similar framework we've extended predicts power law mass profiles on small scales. Hence, it is inconsistent with an Einasto profile.  

It is clear from our analysis that angular momentum plays an essential role in determining the structure of the halo in two important ways. First, the amount of angular momentum at turnaround ($B$) sets the width of the isothermal region. Second, the presence of pericenters softens the inner density slope relative to the FG solution because less mass shells contribute to the enclosed mass. Moreover, the interior density profile is sensitive to the way in which particles are torqued after turnaround ($\varpi$).

If we assume that $\varpi$ is constant for all halos, then this secondary infall model predicts steeper interior density profiles for larger mass halos. More specifically, if we use the value of $\varpi= 0.12$ which gave $\rho\propto r^{-1}$ for galactic size halos, then $\rho\propto r^{-0.66}$ for a $10^8M_{\odot}$ halo and $\rho\propto r^{-1.42}$ for a $10^{15}M_{\odot}$ halo. This trend towards steeper interior slopes for larger mass halos, and hence non-universality, has been noticed in recent numerical simulations \cite{Ricotti,Cen} as well as more general secondary infall models \cite{delpopolo2}. On the other hand, if we assume that all halos have $\rho\propto r^{-1}$ as $r\rightarrow0$, then $\varpi$ must vary with halo mass. More specifically, halos with mass $M< 10^9M_{\odot}$ must have particles which lose angular momentum over time ($\varpi<0$) while halos with mass $M> 10^9M_{\odot}$ must have particles which gain angular momentum over time ($\varpi > 0$). In other words, in order for our self-similar framework to predict universal density profiles, $\varpi$ must conspire to erase any dependence on initial conditions. A more thorough treatment requires the use of N-body simulations, which is beyond the scope of this paper.

It is also possible to predict a dark matter halo's density distribution if one assumes a mapping between a mass shell's initial radius, when the structure is linear, to its final average radius, which is some fraction of its turnaround radius. From this scheme, one can also infer a velocity dispersion, using the virial theorem. Unfortunately, this scheme does not give any information about the halo's velocity anisotropy. Our self-similar prescription discussed above, on the other hand, contains all of the velocity information. Hence, one can reconstruct the velocity anisotropy profile given the trajectory of a mass shell. The velocity anisotropy is significant since it describes to what degree orbits are radial. Moreover, it can break degeneracies between $n$ and $\varpi$ in our halo model. We will discuss the velocity structure of our halo model, including the pseudo-phase-space density profile, in more detail in Paper 2 of this series. There we will once again compare our halo predictions to the recent Aquarius simulation results \cite{Aquarius}.  

While the above self-similar prescription has its clear advantages, it's also unphysical since mass shells at turnaround are radially cold. The same tidal torque mechanisms which cause a tangential velocity dispersion \cite{Hoyle}, should also give rise to a radial velocity dispersion. For a more physical model, one would need to impose self-similarity to a phase space description of the halo and include sources of torque as diffusion terms in the Boltzmann equation. This will be the subject of Paper 3 of this series.    

As we've shown, the way in which particles are torqued after turnaround ($\varpi$) influences the interior power law of the density profile. One way to source this change in angular momentum is through substructure that is aspherically distributed throughout the halo. It is reasonable to assume that substructure dominated by baryons torque halo particles more strongly than substructure dominated by dark matter since baryons can achieve higher densities and hence are not tidally disrupted as easily. If this is the case, then torques sourced by baryons would result in a larger value of $\varpi$ than torques sourced by dark matter. According to the predictions of this secondary infall model, this would lead to less cuspy profiles (See Appendix \ref{sec:ATE} for a more detailed discussion). Therefore a more thorough understanding of $\varpi$ coupled with this simplified model of halo formation could potentially shed light on the Cusp Core problem and thereby possibly bridge the gap between simulations and observations.    

\acknowledgments

The authors acknowledge support from NASA grant NNG06GG99G and thank Robyn Sanderson and Paul Schechter for many useful conversations.

\appendix

\section{Calculating $n_{\rm{eff}}$}
\label{sec:neff}

The effective primordial power spectral index, $n_{\rm{eff}}$, relates our model parameter $n$ to the halo mass $M$. The effective index is defined by:

\begin{equation}
n_{\rm{eff}}\equiv-2\frac{d\ln \sigma_R}{d \ln R}-3
\end{equation}   

\noindent where 

\begin{equation}
\sigma^2_R\equiv\int \frac{d^3k}{(2\pi)^3}P(k)W_R^2(k)
\end{equation}

\noindent and 

\begin{equation}\label{window}
W_R(k)\equiv3\frac{\sin(kR)-(kR)\cos(kR)}{(kR)^3}.
\end{equation}

\noindent The scale $R$ is set by the top hat mass of the halo ($M=4\pi\rho_{m0}R^3/3$), where $\rho_{m0}$ is the dark matter background density today. The power spectrum today, $P(k)$, is given by \cite{takada}:

\begin{equation}\label{power}
P(k)=\left(\frac{2k^2}{5H^2_0\Omega_m}\right)^2P_{\cal{R}}(k)T^2(k)D^2(a=1)
\end{equation}

\noindent where

\begin{equation}
\frac{k^3P_{\cal{R}}(k)}{2\pi^2}=\triangle^2_{\cal{R}}(k_0)\left(\frac{k}{k_0}\right)^{n_s-1},
\end{equation}

\noindent $D$ is the linear growth factor normalized so that $D/a\rightarrow 1$ as $a\rightarrow0$, $a$ is the scale factor, $T(k)$ is the transfer function and we choose cosmological parameters derived from WMAP7: $\triangle^2_{\cal{R}}(k_0)=2.441\times10^{-9}$, $h=0.704$, $\Omega_m=0.272$, $n_s=0.963$, with $k_0=0.002 \;\rm{Mpc}^{-1}$ \cite{wmap}. We calculate $T(k)$ using CMBFAST \cite{cmbfast}.

\section{Tidal Torque Theory}
\label{sec:TT}

We first derive eq.\ (\ref{sigmaWhite}) using cosmological linear perturbation theory. Starting with the Zel'dovich approximation \cite{zeldovich}, we have:

\begin{equation}
\boldsymbol{r}(\boldsymbol{q},t) = a(t)\Big(\boldsymbol{q}-D(t)\boldsymbol{\nabla}\phi(q)\Big)
\end{equation} 

\noindent where $\boldsymbol{r}$ is the physical radius, $\boldsymbol{q}$ is a Lagrangian coordinate, and $\phi$, which is time independent, is related to the Newtonian potential $\Phi$ through the following:

\begin{equation}
\phi = \frac{1}{4\pi G \rho_mDa^2}\Phi
\end{equation}

\noindent where $\rho_m$ is the dark matter background density. The velocity, $\partial\boldsymbol{r}/\partial t$, then is given by:

\begin{equation}
\boldsymbol{v}(\boldsymbol{q},t)=H(t)\boldsymbol{r}(\boldsymbol{q},t)-a(t)\dot{D}(t)\boldsymbol{\nabla}\phi(\boldsymbol{q})
\end{equation}

\noindent where $H\equiv d\ln a/dt$ and dots denote derivatives with respect to time. Therefore, to first order in $\phi$:

\begin{eqnarray}
(\boldsymbol{r}-\boldsymbol{r}_0)\times(\boldsymbol{v}-\boldsymbol{v}_0) = -a^2\dot{D}(\boldsymbol{q}-\boldsymbol{q_0})\times(\boldsymbol{\nabla}\phi-\boldsymbol{\nabla}\phi_0) \nonumber \\
\end{eqnarray}

\noindent where $\boldsymbol{\nabla}\phi_0\equiv\boldsymbol{\nabla}\phi(\boldsymbol{q}_0)$. Plugging this expression into eq.\ (\ref{sigma}), taking an expectation value, defining $\boldsymbol{x}\equiv\boldsymbol{q}-\boldsymbol{q}_0$, rewriting in terms of the velocity perturbation variable $\boldsymbol{\Psi}\equiv -D\boldsymbol{\nabla}\phi$ and using index notation, we find:

\begin{eqnarray} \label{sigma1}
<\tilde{\sigma}^2> &=& \frac{a^4\dot{D}^2}{D^2}\epsilon_{ijk}\epsilon_{ilm}\int_{V_L}d^3x\rho_ma^3x_jx_l \times \nonumber \\
&&<\Big(\Psi_k(0)-\Psi_k(\boldsymbol{x})\Big)\Big(\Psi_m(0)-\Psi_m(\boldsymbol{x})\Big)>_C \nonumber \\
\end{eqnarray}

\noindent where $\epsilon_{ijk}$ is the Levi-Civita tensor and $V_L$ is the Lagrangian volume of the halo. We assume $V_L$ is spherical with radius $x_{\rm{max}}$. Note that to linear order in $\phi$, the Lagrangian coordinate $\boldsymbol{x}$ is equivalent to a comoving coordinate. The subscript $C$ denotes an expectation value taken over constrained gaussian fields since we want to average only over peaks in the density field. 

Using the formalism developed in \cite{bbks}, we choose, for simplicity, to use the zeroth and first order derivatives of the smoothed density field to constrain the velocity perturbation. Our treatment and conventions are identical to that used in Appendix A of Ma and Bertschinger \cite{MaBert}. For more details, please refer to this reference. We find:

\begin{eqnarray}
\left\langle\Psi_i(\boldsymbol{x})\Psi_j(\boldsymbol{x})\right\rangle_C &=& \Big(\sigma^2_{\Psi}-\frac{\bar{\eta}^2(x)}{\sigma^2_1}\Big)(\delta_{ij}-\hat{x}_i\hat{x}_j)+\nonumber \\ \Bigg(\sigma^2_{\Psi}-\frac{x^2\bar{\eta}^2(x)}{\sigma^2_0}&-&\frac{\big(\bar{\xi}(x)-2\bar{\eta}(x)\big)^2}{\sigma^2_1}\Bigg)\hat{x}_i\hat{x}_j  \label{constrained1}\\
\left\langle\Psi_i(\boldsymbol{x})\Psi_j(0)\right\rangle_C &=& \Big(\frac{\gamma(x)}{x}-\frac{\bar{\eta}(0)\bar{\eta}(x)}{\sigma^2_1}\Big)\delta_{ij} + \nonumber \\ \Big(\frac{d\gamma(x)}{dx}-\frac{\gamma(x)}{x}&-&\frac{x\bar{\eta}(0)}{\sigma^2_1}\frac{d\bar{\eta}(x)}{dx}\Big)\hat{x}_i\hat{x}_j \label{Ma} \\
\left\langle\Psi_i(0)\Psi_j(0)\right\rangle_C &=& \Big(\sigma^2_\Psi-\frac{\bar{\eta}^2(0)}{\sigma^2_1}\Big)\delta_{ij} \label{constrained2}
\end{eqnarray}  

\noindent where:

\begin{eqnarray}
\sigma^2_\Psi &\equiv& \frac{1}{3}\int\frac{d^3k}{(2\pi)^3}k^{-2}P(k) \\
\sigma^2_0&\equiv& \int\frac{d^3k}{(2\pi)^3}P(k)W^2_R(k) \\
\sigma^2_1&\equiv& \frac{1}{3}\int\frac{d^3k}{(2\pi)^3}k^2P(k)W^2_R(k) \\
\bar{\eta}(x)&\equiv& \int\frac{d^3k}{(2\pi)^3}P(k)W_R(k)\frac{j_1(kx)}{kx} \\
\bar{\xi}(x)&\equiv& \int\frac{d^3k}{(2\pi)^3}P(k)W_R(k)j_0(kx) \\
\gamma(x)&\equiv& \int\frac{d^3k}{(2\pi)^3}P(k)k^{-3}j_1(kx) .
\end{eqnarray}

\noindent $W_R(k)$ and $P(k)$ are defined in eq.\ (\ref{window}) and (\ref{power}) respectively and the spherical Bessel functions are $j_0(x)=x^{-1}\sin x$ and $j_1(x)=x^{-2}(\sin x-x\cos x)$. Eq.\ (\ref{Ma}) corrects an error in equation (A15) of Ref. \cite{MaBert}. 

Notice from eqs.\  (\ref{constrained1}) through (\ref{constrained2}) that the expressions separate into terms proportional to $\delta_{ij}$ and terms proportional to $\hat{x}_i\hat{x}_j$. The terms proportional to $\hat{x}_i\hat{x}_j$ vanish in eq.\ (\ref{sigma1}) because of the antisymmetry of the Levi-Civita tensors. Eq.\ (\ref{sigma1}) then reduces to:

\begin{eqnarray}\label{tildesigma1}
\left\langle\tilde{\sigma}^2\right\rangle_M=2\frac{a^4\dot{D}^2}{D^2}\int_{V_L}d^3x\rho_ma^3x^2f(x,R)
\end{eqnarray}

\noindent where 

\begin{eqnarray}
f(x,R)= 2\sigma^2_{\Psi}-\frac{2\gamma(x)}{x}-\frac{\bar{\eta}^2(x)}{\sigma^2_1}+\frac{2\bar{\eta}(0)\bar{\eta}(x)}{\sigma^2_1}-\frac{\bar{\eta}^2(0)}{\sigma^2_1}\nonumber \\
\end{eqnarray}

\noindent Last, defining $u=x/x_{\rm{max}}$, we find:

\begin{eqnarray}
\left\langle\tilde{\sigma}^2\right\rangle_M &=& 6\frac{a^4\dot{D}^2}{D^2}Mx^2_{max}\int^1_0u^4f(ux_{\rm{max}},R)du \nonumber \\
&\equiv& 6a^4\dot{D}^2 Mx^2_{\rm{max}}A^2(R)
\end{eqnarray} 

\noindent In the above $A$ has units of Mpc and $x_{\rm{max}}=R$ since the scale of the galaxy today is equivalent to its lagrangian size to linear order in perturbation theory. Note that $f/D^2$ is time independent. 

Now, we derive eq.\ (\ref{pBconstraints}). Consistency with the secondary infall model demands that we assume $\Omega_m=1$. Equating the time dependence of eq.\ (\ref{sigmaWhite}) and eq.\ (\ref{sigmaModel}) for an Einstein de-Sitter universe ($D=a$) at early times ($r_{\rm{max}}\propto t^{2/3}$), we find $p=2n$ where $p=2\gamma+4$. Given this relationship, we now equate eq.\ (\ref{sigmaWhite}) to eq.\ (\ref{sigmaModel}) and solve for $B$. We find:
 
\begin{eqnarray}\label{B1}
B=\frac{2}{3}\sqrt{2(7-2n)}a^2\left(\frac{r_{\max}}{r_{\rm{ta}}}\right)^{n-1}\frac{A(R)}{r_{\rm{ta}}}
\end{eqnarray}

\noindent where we've used eqs.\ (\ref{MDef}) and (\ref{Minit}) evaluated at early times to cancel the mass as well as the relationship $r_{\rm{max}}=ax_{\rm{max}}$. Now we evaluate eq.\ (\ref{B1}) at early times since tidal torque theory only applies when the halo is linear. To relate $r_{\rm{max}}$ at some initial time ($t_i$) to $r_{\rm{ta}}$ today ($t_0$), we use the conservation of mass.

\begin{equation}\label{Rrelation}
\frac{4\pi}{3}\rho_{B}(t_i)r^3_{\rm{max}}(t_i)=\frac{4\pi}{3}{\cal{M}}(1)\rho_{B}(t_0)r^3_{ta}(t_0)
\end{equation}

\noindent Evaluating eq.\ (\ref{B1}) at $t_i$ with the use of eq.\ (\ref{Rrelation}) and noting that $r_{\rm{ta}}\propto t^{\beta}$, we find:  

\begin{eqnarray}
B=\frac{2}{3}\sqrt{2(7-2n)}{\cal{M}}(1)^{(n-1)/3}\frac{A(R)}{r_{\rm{ta}}(t_0)}
\end{eqnarray}

\noindent As expected, the time dependence of $B$ vanishes. Last, assuming $r_{\rm{ta}}(t_0)=R$, we reproduce eq.\ (\ref{pBconstraints}). The quantities $n,R,A(R)$ are calculated in an $\Omega_m=1$ universe with the same background matter density and power spectrum of $\Lambda$CDM today. This ensures that the statistics, mass, and size of halos in both universes are equivalent today. 

As mentioned previously, eq.\ (\ref{pBconstraints}) overestimates the angular momentum of particles at turnaround by a factor of 1.5 to 2.3. One potential source of error is to assume the lagrangian volume is spherical. Assuming an ellipsoidal lagrangian volume with axis ratios $1:a:b$, we find that $B$ is at most reduced by 8$\%$ when $0.5<a,b<1$. The discrepancy may be caused by not including higher order constraints on the smoothed density field. However, comparing $B$ when calculated using zeroth and first order derivative constraints with $B$ when calculated without using constraints results in only percent level differences; so it seems unlikely that constraints would have a significant effect. 

N-body simulations use friends-of-friends group finders in order to identify halos and subhalos \cite{Davis,Aquarius2}. This algorithm, however, removes particles that are grouped to neighboring halos and hence neglects a contribution to $\tilde{\sigma}^2$. Trying to mimick this selection effect, we replaced eq.\ (\ref{tildesigma1}) with 

\begin{equation}
\left\langle\tilde{\sigma}^2\right\rangle_M=2\frac{a^4\dot{D}^2}{D^2}\int^{\infty}_{0}d^3xe^{-x^2/2\bar{R}^2}\rho_ma^3x^2f(x,R)
\end{equation}

\noindent where $\bar{R}\equiv(2/9\pi)^{1/6}R$ ensures that the mass enclosed within the lagrangian volume is equal to the mass of the halo calculated by the simulation. However, calculating $B$ in this manner leads to overestimating the angular momentum at turnaround by $\sim4$, as opposed to $\sim2$ beforehand. Since $f(x,R)$ is an increasing function of $x$ ($f\sim x^{1.34}$ near $R$ for $10^{12}M_{\odot}$ halos), most of the contribution to $A(R)$ comes from close to $R$. Therefore, while the gaussian cutoff decreases the contribution to $B$ around $R$, it includes contributions beyond $R$, leading to a worse estimate. This highlights that $B$ significantly depends on the outer parts of the halo. Hence, overestimating $B$ by $\sim2$ is reasonable.  

Last, note that the parameter $B$ is set during the linear regime. Assuming that the shell is dominated by substructure at turnaround, nonlinear interactions like dynamical friction and tidal stripping play an important role from the time of turnaround to the first pericenter passage \cite{gan,zentner}. As time goes on, these effects become less important since substructure in the shell becomes subdominant. Including these extra interactions should lead to smaller estimates of $B$ at first pericenter passage and hence potentially explain our overestimate, but is beyond the scope of this work.  

\section{Evolution After Turnaround}
\label{sec:ATE}

In this Appendix, we use dimensional analysis in order to gain intuition about $\varpi$, a parameter that describes tidal torque after turnaround. First, consider the time derivative of $L^2$, where $\boldsymbol{L}$ is the angular momentum per unit mass of a particle with radius $r$ at time $t$.

\begin{equation}\label{L2}
\frac{dL^2}{dt} = 2 \Big(r^2(\boldsymbol{v}\cdot\boldsymbol{a})-(\boldsymbol{r}\cdot\boldsymbol{a})(\boldsymbol{r}\cdot\boldsymbol{v})\Big)
\end{equation}

\noindent In the above, $\boldsymbol{a}$ ($\boldsymbol{v}$) is the acceleration (velocity) of the particle. We now decompose the acceleration vector into a radial ($\hat{\boldsymbol{r}}$) and tangential ($\hat{\boldsymbol{t}}$) component and use this basis to rewrite the velocity vector. 

\begin{eqnarray}
\boldsymbol{a}&=&a_r\hat{\boldsymbol{r}}+a_t\hat{\boldsymbol{t}} \\
\boldsymbol{v}&=&v_r\hat{\boldsymbol{r}}+v_t\hat{\boldsymbol{t}}+v_p\hat{\boldsymbol{p}}
\end{eqnarray}

\noindent The direction $\hat{\boldsymbol{p}}$ is orthogonal to both $\hat{\boldsymbol{r}}$ and $\hat{\boldsymbol{t}}$. Note that all basis vectors depend on position. Plugging in the above decomposed vectors into eq.\ (\ref{L2}), we find:

\begin{equation}\label{L2simple}
\frac{dL^2}{dt}=2r^2v_ta_t.
\end{equation}

As expected, changes in $L^2$ are sourced by deviations from spherical symmetry that create nonzero $\boldsymbol{a}_t$. Now, imagine a spherically symmetric halo, roughly described by our self-similar infall model with $\varpi>0$, with a clump of mass $m$ in the shell at radius $r_2$. We assume that $m$ is small enough so that it does not influence the radial equation of motion of the shell at $r$. We focus on $\varpi>0$ since this is required in order for the density profile of a $10^{12}M_{\odot}$ halo to be consistent with the NFW profile (Section \ref{nbody}). 

Next, consider averaging $dL^2/dt$ over an orbital period and over a spherical shell of radius $r$, in order to compare the change in angular momentum sourced by the clump at $r_2$ to our model's prescription for angular momentum evolution. For $\varpi>0$, orbits are roughly circular at late times. Hence, we assume averaging over an orbital period is equivalent to evaluating the right hand side of eq.\ (\ref{L2simple}) at roughly the apocenter radius of the shell. As described in Section \ref{sec:SS}, the orbital planes of all particles in a shell at $r$ are randomly aligned. Therefore we expect $v_t$ averaged over a sphere to vanish. However, if there exists an excess mass $m$, then all the particles will be pulled slightly in that direction, leading to a nonzero average. We therefore assume $v_t\propto Pa_t$ where $P$, the orbital period of the particle, is taken to be a dynamical time ($P\propto\rho(r)^{-1/2}$). Last assuming $r\ll r_2$, we find:

\begin{equation}
\left\langle\frac{dL^2}{dt}\right\rangle\propto\;\;\frac{r^2m^2}{r^4_2\rho^{1/2}}.
\end{equation}

In order for the secondary infall halo model to be consistent, the right hand and left hand side must have the same time scaling. Assuming $m\propto t^{\mu}$, associating $r$ and $r_2$ with their respective apocenters, and noting from eq.\ (\ref{relations3}) that $r\propto t^{2\varpi}$, $r_2\propto t^{2\varpi}$, and $\rho\propto t^{-6\varpi}$, we find:

\begin{equation}\label{varpiConst}
\varpi=\frac{1}{3}(1+2\mu)
\end{equation} 

The same scaling relationship holds for $r\gg r_2$. Hence one could imagine substructure in the shell at $r$ sourcing a change in $L^2$ of the shell at $r_2$ and substructure in the shell at $r_2$ sourcing a change in $L^2$ of the shell at $r$. Therefore, a hierarchy of substructure non-spherically distributed, which is subdominant to the monopole contribution of the halo, would result in a halo roughly consistent with our described secondary infall model.  

Eq.\ (\ref{varpiConst}), which is only valid for $\mu>-1/2$ since we assumed $\varpi>0$, together with eq.\ (\ref{relations3}) relates the steepness of the inner density profile to the mass loss rate of substructure. If the clump does not lose mass ($\mu=0$), then $\varpi=1/3$ implies $\rho\propto r^0$. If the clump loses mass ($\mu<0$), eq.\ (\ref{relations3}) predicts steeper density profiles. Substructure dominated by baryons will lose less mass than substructure dominated by dark matter, since baryons clump more easily and hence have higher densities. Therefore, according to the above analysis, pure dark matter simulations {\it{should}} have steeper density profiles than galaxies which include baryons. This is expected since baryons stir particles around more efficiently, causing larger pericenters and less dense interiors. A more thorough treatment that involves constraining $\varpi$ with simulations is beyond the scope of this paper. 

\bibliography{SSpaper}

%Merlin.mbs v4.21 2009-07-09.
\begin{thebibliography}{10}%
\makeatletter
\providecommand \@ifxundefined [1]{%
 \ifx #1\undefined \expandafter \@firstoftwo
 \else \expandafter \@secondoftwo
\fi
}%
\providecommand \@ifnum [1]{%
 \ifnum #1\expandafter \@firstoftwo
 \else \expandafter \@secondoftwo
\fi
}%
\providecommand \enquote [1]{``#1''}%
\providecommand \bibnamefont  [1]{#1}%
\providecommand \bibfnamefont [1]{#1}%
\providecommand \citenamefont [1]{#1}%
\providecommand\href[0]{\@sanitize\@href}%
\providecommand\@href[1]{\endgroup\@@startlink{#1}\endgroup\@@href}%
\providecommand\@@href[1]{#1\@@endlink}%
\providecommand \@sanitize [0]{\begingroup\catcode`\&12\catcode`\#12\relax}%
\@ifxundefined \pdfoutput {\@firstoftwo}{%
 \@ifnum{\z@=\pdfoutput}{\@firstoftwo}{\@secondoftwo}%
}{%
 \providecommand\@@startlink[1]{\leavevmode\special{html:<a href="#1">}}%
 \providecommand\@@endlink[0]{\special{html:</a>}}%
}{%
 \providecommand\@@startlink[1]{%
  \leavevmode
  \pdfstartlink
   attr{/Border[0 0 1 ]/H/I/C[0 1 1]}%
   user{/Subtype/Link/A<</Type/Action/S/URI/URI(#1)>>}%
  \relax
 }%
 \providecommand\@@endlink[0]{\pdfendlink}%
}%
\providecommand \url  [0]{\begingroup\@sanitize \@url }%
\providecommand \@url [1]{\endgroup\@href {#1}{\urlprefix}}%
\providecommand \urlprefix [0]{URL }%
\providecommand \Eprint[0]{\href }%
\@ifxundefined \urlstyle {%
  \providecommand \doi [1]{doi:\discretionary{}{}{}#1}%
}{%
  \providecommand \doi [0]{doi:\discretionary{}{}{}\begingroup
  \urlstyle{rm}\Url }%
}%
\providecommand \doibase [0]{http://dx.doi.org/}%
\providecommand \Doi[1]{\href{\doibase#1}}%
\providecommand \bibAnnote [3]{%
  \BibitemShut{#1}%
  \begin{quotation}\noindent
    \textsc{Key:}\ #2\\\textsc{Annotation:}\ #3%
  \end{quotation}%
}%
\providecommand \bibAnnoteFile [2]{%
  \IfFileExists{#2}{\bibAnnote {#1} {#2} {\input{#2}}}{}%
}%
\providecommand \typeout [0]{\immediate \write \m@ne }%
\providecommand \selectlanguage [0]{\@gobble}%
\providecommand \bibinfo [0]{\@secondoftwo}%
\providecommand \bibfield [0]{\@secondoftwo}%
\providecommand \translation [1]{[#1]}%
\providecommand \BibitemOpen[0]{}%
\providecommand \bibitemStop [0]{}%
\providecommand \bibitemNoStop [0]{.\EOS\space}%
\providecommand \EOS [0]{\spacefactor3000\relax}%
\providecommand \BibitemShut [1]{\csname bibitem#1\endcsname}%
%</preamble>
\bibitem{GunnGott72}%
  \BibitemOpen
  \bibfield{author}{%
  \bibinfo {author} {\bibfnamefont{J.~E.}\ \bibnamefont{{Gunn}}}\ and\ \bibinfo
  {author} {\bibfnamefont{J.~R.}\ \bibnamefont{{Gott}}, \bibfnamefont{III}},\
  }%
  \bibfield{journal}{%
  \Doi{10.1086/151605}{\bibinfo {journal} {\apj}}\ }%
  \textbf{\bibinfo {volume} {176}},\ \bibinfo {pages} {1} (\bibinfo {month}
  {Aug.}\ \bibinfo {year} {1972})%
  \bibAnnoteFile{NoStop}{GunnGott72}%
\bibitem{Gott75}%
  \BibitemOpen
  \bibfield{author}{%
  \bibinfo {author} {\bibfnamefont{J.~R.}\ \bibnamefont{{Gott}},
  \bibfnamefont{III}},\ }%
  \bibfield{journal}{%
  \Doi{10.1086/153887}{\bibinfo {journal} {\apj}}\ }%
  \textbf{\bibinfo {volume} {201}},\ \bibinfo {pages} {296} (\bibinfo {month}
  {Oct.}\ \bibinfo {year} {1975})%
  \bibAnnoteFile{NoStop}{Gott75}%
\bibitem{Gunn77}%
  \BibitemOpen
  \bibfield{author}{%
  \bibinfo {author} {\bibfnamefont{J.~E.}\ \bibnamefont{{Gunn}}},\ }%
  \bibfield{journal}{%
  \Doi{10.1086/155715}{\bibinfo {journal} {\apj}}\ }%
  \textbf{\bibinfo {volume} {218}},\ \bibinfo {pages} {592} (\bibinfo {month}
  {Dec.}\ \bibinfo {year} {1977})%
  \bibAnnoteFile{NoStop}{Gunn77}%
\bibitem{Bert}%
  \BibitemOpen
  \bibfield{author}{%
  \bibinfo {author} {\bibfnamefont{E.}~\bibnamefont{{Bertschinger}}},\ }%
  \bibfield{journal}{%
  \Doi{10.1086/191028}{\bibinfo {journal} {\apj}}\ }%
  \textbf{\bibinfo {volume} {58}},\ \bibinfo {pages} {39} (\bibinfo {month}
  {May}\ \bibinfo {year} {1985})%
  \bibAnnoteFile{NoStop}{Bert}%
\bibitem{FG}%
  \BibitemOpen
  \bibfield{author}{%
  \bibinfo {author} {\bibfnamefont{J.~A.}\ \bibnamefont{{Fillmore}}}\ and\
  \bibinfo {author} {\bibfnamefont{P.}~\bibnamefont{{Goldreich}}},\ }%
  \bibfield{journal}{%
  \Doi{10.1086/162070}{\bibinfo {journal} {\apj}}\ }%
  \textbf{\bibinfo {volume} {281}},\ \bibinfo {pages} {1} (\bibinfo {month}
  {Jun.}\ \bibinfo {year} {1984})%
  \bibAnnoteFile{NoStop}{FG}%
\bibitem{Aquarius}%
  \BibitemOpen
  \bibfield{author}{%
  \bibinfo {author} {\bibfnamefont{J.~F.}\ \bibnamefont{{Navarro}}}, \bibinfo
  {author} {\bibfnamefont{A.}~\bibnamefont{{Ludlow}}}, \bibinfo {author}
  {\bibfnamefont{V.}~\bibnamefont{{Springel}}}, \bibinfo {author}
  {\bibfnamefont{J.}~\bibnamefont{{Wang}}}, \bibinfo {author}
  {\bibfnamefont{M.}~\bibnamefont{{Vogelsberger}}}, \bibinfo {author}
  {\bibfnamefont{S.~D.~M.}\ \bibnamefont{{White}}}, \bibinfo {author}
  {\bibfnamefont{A.}~\bibnamefont{{Jenkins}}}, \bibinfo {author}
  {\bibfnamefont{C.~S.}\ \bibnamefont{{Frenk}}},\ and\ \bibinfo {author}
  {\bibfnamefont{A.}~\bibnamefont{{Helmi}}},\ }%
  \bibfield{journal}{%
  \Doi{10.1111/j.1365-2966.2009.15878.x}{\bibinfo {journal} {Mon. Not. R.
  Astron. Soc.}}\ }%
  \textbf{\bibinfo {volume} {402}},\ \bibinfo {pages} {21} (\bibinfo {month}
  {Feb.}\ \bibinfo {year} {2010}),\
  \Eprint{http://arxiv.org/abs/0810.1522}{arXiv:0810.1522}%
  \bibAnnoteFile{NoStop}{Aquarius}%
\bibitem{Graham}%
  \BibitemOpen
  \bibfield{author}{%
  \bibinfo {author} {\bibfnamefont{A.~W.}\ \bibnamefont{{Graham}}}, \bibinfo
  {author} {\bibfnamefont{D.}~\bibnamefont{{Merritt}}}, \bibinfo {author}
  {\bibfnamefont{B.}~\bibnamefont{{Moore}}}, \bibinfo {author}
  {\bibfnamefont{J.}~\bibnamefont{{Diemand}}},\ and\ \bibinfo {author}
  {\bibfnamefont{B.}~\bibnamefont{{Terzi{\'c}}}},\ }%
  \bibfield{journal}{%
  \Doi{10.1086/508990}{\bibinfo {journal} {Astron. J.}}\ }%
  \textbf{\bibinfo {volume} {132}},\ \bibinfo {pages} {2701} (\bibinfo {month}
  {Dec.}\ \bibinfo {year} {2006}),\
  \Eprint{http://arxiv.org/abs/arXiv:astro-ph/0608613}{arXiv:astro-ph/0608613}%
  \bibAnnoteFile{NoStop}{Graham}%
\bibitem{Lactea}%
  \BibitemOpen
  \bibfield{author}{%
  \bibinfo {author} {\bibfnamefont{J.}~\bibnamefont{{Diemand}}}, \bibinfo
  {author} {\bibfnamefont{B.}~\bibnamefont{{Moore}}},\ and\ \bibinfo {author}
  {\bibfnamefont{J.}~\bibnamefont{{Stadel}}},\ }%
  \bibfield{journal}{%
  \Doi{10.1111/j.1365-2966.2004.08094.x}{\bibinfo {journal} {Mon. Not. R.
  Astron. Soc.}}\ }%
  \textbf{\bibinfo {volume} {353}},\ \bibinfo {pages} {624} (\bibinfo {month}
  {Sep.}\ \bibinfo {year} {2004}),\
  \Eprint{http://arxiv.org/abs/arXiv:astro-ph/0402267}{arXiv:astro-ph/0402267}%
  \bibAnnoteFile{NoStop}{Lactea}%
\bibitem{deBlok03}%
  \BibitemOpen
  \bibfield{author}{%
  \bibinfo {author} {\bibfnamefont{W.~J.~G.}\ \bibnamefont{{de Blok}}},\ }%
  in\ \emph{\bibinfo {booktitle} {Revista Mexicana de Astronomia y Astrofisica
  Conference Series}},\ \bibinfo {series} {Revista Mexicana de Astronomia y
  Astrofisica, vol. 27}, Vol.~\bibinfo {volume} {17},\ \bibinfo {editor}
  {edited by\ \bibinfo {editor} {\bibnamefont{{V.~Avila-Reese, C.~Firmani,
  C.~S.~Frenk, \& C.~Allen}}}}\ (\bibinfo {year} {2003})\ pp.\ \bibinfo {pages}
  {17--18}%
  \bibAnnoteFile{NoStop}{deBlok03}%
\bibitem{Salucci1}%
  \BibitemOpen
  \bibfield{author}{%
  \bibinfo {author} {\bibfnamefont{G.}~\bibnamefont{{Gentile}}}, \bibinfo
  {author} {\bibfnamefont{P.}~\bibnamefont{{Salucci}}}, \bibinfo {author}
  {\bibfnamefont{U.}~\bibnamefont{{Klein}}}, \bibinfo {author}
  {\bibfnamefont{D.}~\bibnamefont{{Vergani}}},\ and\ \bibinfo {author}
  {\bibfnamefont{P.}~\bibnamefont{{Kalberla}}},\ }%
  \bibfield{journal}{%
  \Doi{10.1111/j.1365-2966.2004.07836.x}{\bibinfo {journal} {Mon. Not. R.
  Astron. Soc.}}\ }%
  \textbf{\bibinfo {volume} {351}},\ \bibinfo {pages} {903} (\bibinfo {month}
  {Jul.}\ \bibinfo {year} {2004}),\
  \Eprint{http://arxiv.org/abs/arXiv:astro-ph/0403154}{arXiv:astro-ph/0403154}%
  \bibAnnoteFile{NoStop}{Salucci1}%
\bibitem{Salucci2}%
  \BibitemOpen
  \bibfield{author}{%
  \bibinfo {author} {\bibfnamefont{P.}~\bibnamefont{{Salucci}}}, \bibinfo
  {author} {\bibfnamefont{A.}~\bibnamefont{{Lapi}}}, \bibinfo {author}
  {\bibfnamefont{C.}~\bibnamefont{{Tonini}}}, \bibinfo {author}
  {\bibfnamefont{G.}~\bibnamefont{{Gentile}}}, \bibinfo {author}
  {\bibfnamefont{I.}~\bibnamefont{{Yegorova}}},\ and\ \bibinfo {author}
  {\bibfnamefont{U.}~\bibnamefont{{Klein}}},\ }%
  \bibfield{journal}{%
  \Doi{10.1111/j.1365-2966.2007.11696.x}{\bibinfo {journal} {Mon. Not. R.
  Astron. Soc.}}\ }%
  \textbf{\bibinfo {volume} {378}},\ \bibinfo {pages} {41} (\bibinfo {month}
  {Jun.}\ \bibinfo {year} {2007}),\
  \Eprint{http://arxiv.org/abs/arXiv:astro-ph/0703115}{arXiv:astro-ph/0703115}%
  \bibAnnoteFile{NoStop}{Salucci2}%
\bibitem{Salucci3}%
  \BibitemOpen
  \bibfield{author}{%
  \bibinfo {author} {\bibfnamefont{F.}~\bibnamefont{{Donato}}}, \bibinfo
  {author} {\bibfnamefont{G.}~\bibnamefont{{Gentile}}}, \bibinfo {author}
  {\bibfnamefont{P.}~\bibnamefont{{Salucci}}}, \bibinfo {author}
  {\bibfnamefont{C.}~\bibnamefont{{Frigerio Martins}}}, \bibinfo {author}
  {\bibfnamefont{M.~I.}\ \bibnamefont{{Wilkinson}}}, \bibinfo {author}
  {\bibfnamefont{G.}~\bibnamefont{{Gilmore}}}, \bibinfo {author}
  {\bibfnamefont{E.~K.}\ \bibnamefont{{Grebel}}}, \bibinfo {author}
  {\bibfnamefont{A.}~\bibnamefont{{Koch}}},\ and\ \bibinfo {author}
  {\bibfnamefont{R.}~\bibnamefont{{Wyse}}},\ }%
  \bibfield{journal}{%
  \Doi{10.1111/j.1365-2966.2009.15004.x}{\bibinfo {journal} {Mon. Not. R.
  Astron. Soc.}}\ }%
  \textbf{\bibinfo {volume} {397}},\ \bibinfo {pages} {1169} (\bibinfo {month}
  {Aug.}\ \bibinfo {year} {2009}),\
  \Eprint{http://arxiv.org/abs/0904.4054}{arXiv:0904.4054 [astro-ph.CO]}%
  \bibAnnoteFile{NoStop}{Salucci3}%
\bibitem{RydenGunn}%
  \BibitemOpen
  \bibfield{author}{%
  \bibinfo {author} {\bibfnamefont{B.~S.}\ \bibnamefont{{Ryden}}}\ and\
  \bibinfo {author} {\bibfnamefont{J.~E.}\ \bibnamefont{{Gunn}}},\ }%
  \bibfield{journal}{%
  \Doi{10.1086/165349}{\bibinfo {journal} {\apj}}\ }%
  \textbf{\bibinfo {volume} {318}},\ \bibinfo {pages} {15} (\bibinfo {month}
  {Jul.}\ \bibinfo {year} {1987})%
  \bibAnnoteFile{NoStop}{RydenGunn}%
\bibitem{Nusser}%
  \BibitemOpen
  \bibfield{author}{%
  \bibinfo {author} {\bibfnamefont{A.}~\bibnamefont{{Nusser}}},\ }%
  \bibfield{journal}{%
  \Doi{10.1046/j.1365-8711.2001.04527.x}{\bibinfo {journal} {Mon. Not. R.
  Astron. Soc.}}\ }%
  \textbf{\bibinfo {volume} {325}},\ \bibinfo {pages} {1397} (\bibinfo {month}
  {Aug.}\ \bibinfo {year} {2001}),\
  \Eprint{http://arxiv.org/abs/arXiv:astro-ph/0008217}{arXiv:astro-ph/0008217}%
  \bibAnnoteFile{NoStop}{Nusser}%
\bibitem{Hiotelis}%
  \BibitemOpen
  \bibfield{author}{%
  \bibinfo {author} {\bibfnamefont{N.}~\bibnamefont{{Hiotelis}}},\ }%
  \bibfield{journal}{%
  \Doi{10.1051/0004-6361:20011620}{\bibinfo {journal} {Astron. Astrophys.}}\ }%
  \textbf{\bibinfo {volume} {382}},\ \bibinfo {pages} {84} (\bibinfo {month}
  {Jan.}\ \bibinfo {year} {2002}),\
  \Eprint{http://arxiv.org/abs/arXiv:astro-ph/0111324}{arXiv:astro-ph/0111324}%
  \bibAnnoteFile{NoStop}{Hiotelis}%
\bibitem{WilliamsEtAl}%
  \BibitemOpen
  \bibfield{author}{%
  \bibinfo {author} {\bibfnamefont{L.~L.~R.}\ \bibnamefont{{Williams}}},
  \bibinfo {author} {\bibfnamefont{A.}~\bibnamefont{{Babul}}},\ and\ \bibinfo
  {author} {\bibfnamefont{J.~J.}\ \bibnamefont{{Dalcanton}}},\ }%
  \bibfield{journal}{%
  \Doi{10.1086/381722}{\bibinfo {journal} {\apj}}\ }%
  \textbf{\bibinfo {volume} {604}},\ \bibinfo {pages} {18} (\bibinfo {month}
  {Mar.}\ \bibinfo {year} {2004}),\
  \Eprint{http://arxiv.org/abs/arXiv:astro-ph/0312002}{arXiv:astro-ph/0312002}%
  \bibAnnoteFile{NoStop}{WilliamsEtAl}%
\bibitem{Sikivie}%
  \BibitemOpen
  \bibfield{author}{%
  \bibinfo {author} {\bibfnamefont{P.}~\bibnamefont{{Sikivie}}}, \bibinfo
  {author} {\bibfnamefont{I.~I.}\ \bibnamefont{{Tkachev}}},\ and\ \bibinfo
  {author} {\bibfnamefont{Y.}~\bibnamefont{{Wang}}},\ }%
  \bibfield{journal}{%
  \Doi{10.1103/PhysRevD.56.1863}{\bibinfo {journal} {\prd}}\ }%
  \textbf{\bibinfo {volume} {56}},\ \bibinfo {pages} {1863} (\bibinfo {month}
  {Aug.}\ \bibinfo {year} {1997}),\
  \Eprint{http://arxiv.org/abs/arXiv:astro-ph/9609022}{arXiv:astro-ph/9609022}%
  \bibAnnoteFile{NoStop}{Sikivie}%
\bibitem{DelPopolo}%
  \BibitemOpen
  \bibfield{author}{%
  \bibinfo {author} {\bibfnamefont{A.}~\bibnamefont{{Del Popolo}}},\ }%
  \bibfield{journal}{%
  \Doi{10.1088/0004-637X/698/2/2093}{\bibinfo {journal} {\apj}}\ }%
  \textbf{\bibinfo {volume} {698}},\ \bibinfo {pages} {2093} (\bibinfo {month}
  {Jun.}\ \bibinfo {year} {2009}),\
  \Eprint{http://arxiv.org/abs/0906.4447}{arXiv:0906.4447}%
  \bibAnnoteFile{NoStop}{DelPopolo}%
\bibitem{WhiteZaritsky}%
  \BibitemOpen
  \bibfield{author}{%
  \bibinfo {author} {\bibfnamefont{S.~D.~M.}\ \bibnamefont{{White}}}\ and\
  \bibinfo {author} {\bibfnamefont{D.}~\bibnamefont{{Zaritsky}}},\ }%
  \bibfield{journal}{%
  \Doi{10.1086/171552}{\bibinfo {journal} {\apj}}\ }%
  \textbf{\bibinfo {volume} {394}},\ \bibinfo {pages} {1} (\bibinfo {month}
  {Jul.}\ \bibinfo {year} {1992})%
  \bibAnnoteFile{NoStop}{WhiteZaritsky}%
\bibitem{LeDelliou}%
  \BibitemOpen
  \bibfield{author}{%
  \bibinfo {author} {\bibfnamefont{M.}~\bibnamefont{{Le Delliou}}}\ and\
  \bibinfo {author} {\bibfnamefont{R.~N.}\ \bibnamefont{{Henriksen}}},\ }%
  \bibfield{journal}{%
  \Doi{10.1051/0004-6361:20030922}{\bibinfo {journal} {Astron. Astrophys.}}\ }%
  \textbf{\bibinfo {volume} {408}},\ \bibinfo {pages} {27} (\bibinfo {month}
  {Sep.}\ \bibinfo {year} {2003}),\
  \Eprint{http://arxiv.org/abs/arXiv:astro-ph/0307046}{arXiv:astro-ph/0307046}%
  \bibAnnoteFile{NoStop}{LeDelliou}%
\bibitem{Ascasibar}%
  \BibitemOpen
  \bibfield{author}{%
  \bibinfo {author} {\bibfnamefont{Y.}~\bibnamefont{{Ascasibar}}}, \bibinfo
  {author} {\bibfnamefont{G.}~\bibnamefont{{Yepes}}}, \bibinfo {author}
  {\bibfnamefont{S.}~\bibnamefont{{Gottl{\"o}ber}}},\ and\ \bibinfo {author}
  {\bibfnamefont{V.}~\bibnamefont{{M{\"u}ller}}},\ }%
  \bibfield{journal}{%
  \Doi{10.1111/j.1365-2966.2004.08005.x}{\bibinfo {journal} {Mon. Not. R.
  Astron. Soc.}}\ }%
  \textbf{\bibinfo {volume} {352}},\ \bibinfo {pages} {1109} (\bibinfo {month}
  {Aug.}\ \bibinfo {year} {2004}),\
  \Eprint{http://arxiv.org/abs/arXiv:astro-ph/0312221}{arXiv:astro-ph/0312221}%
  \bibAnnoteFile{NoStop}{Ascasibar}%
\bibitem{HoffmanShaham}%
  \BibitemOpen
  \bibfield{author}{%
  \bibinfo {author} {\bibfnamefont{Y.}~\bibnamefont{{Hoffman}}}\ and\ \bibinfo
  {author} {\bibfnamefont{J.}~\bibnamefont{{Shaham}}},\ }%
  \bibfield{journal}{%
  \Doi{10.1086/163498}{\bibinfo {journal} {\apj}}\ }%
  \textbf{\bibinfo {volume} {297}},\ \bibinfo {pages} {16} (\bibinfo {month}
  {Oct.}\ \bibinfo {year} {1985})%
  \bibAnnoteFile{NoStop}{HoffmanShaham}%
\bibitem{Hoyle}%
  \BibitemOpen
  \bibfield{author}{%
  \bibinfo {author} {\bibfnamefont{F.}~\bibnamefont{{Hoyle}}},\ }%
  in\ \emph{\bibinfo {booktitle} {Problems of Cosmical Aerodynamics}}\
  (\bibinfo {year} {1951})\ pp.\ \bibinfo {pages} {195--197}%
  \bibAnnoteFile{NoStop}{Hoyle}%
\bibitem{Hayashi}%
  \BibitemOpen
  \bibfield{author}{%
  \bibinfo {author} {\bibfnamefont{E.}~\bibnamefont{{Hayashi}}}, \bibinfo
  {author} {\bibfnamefont{J.~F.}\ \bibnamefont{{Navarro}}},\ and\ \bibinfo
  {author} {\bibfnamefont{V.}~\bibnamefont{{Springel}}},\ }%
  \bibfield{journal}{%
  \Doi{10.1111/j.1365-2966.2007.11599.x}{\bibinfo {journal} {Mon. Not. R.
  Astron. Soc.}}\ }%
  \textbf{\bibinfo {volume} {377}},\ \bibinfo {pages} {50} (\bibinfo {month}
  {May}\ \bibinfo {year} {2007}),\
  \Eprint{http://arxiv.org/abs/arXiv:astro-ph/0612327}{arXiv:astro-ph/0612327}%
  \bibAnnoteFile{NoStop}{Hayashi}%
\bibitem{Spin1}%
  \BibitemOpen
  \bibfield{author}{%
  \bibinfo {author} {\bibfnamefont{J.}~\bibnamefont{{Barnes}}}\ and\ \bibinfo
  {author} {\bibfnamefont{G.}~\bibnamefont{{Efstathiou}}},\ }%
  \bibfield{journal}{%
  \Doi{10.1086/165480}{\bibinfo {journal} {\apj}}\ }%
  \textbf{\bibinfo {volume} {319}},\ \bibinfo {pages} {575} (\bibinfo {month}
  {Aug.}\ \bibinfo {year} {1987})%
  \bibAnnoteFile{NoStop}{Spin1}%
\bibitem{Spin2}%
  \BibitemOpen
  \bibfield{author}{%
  \bibinfo {author} {\bibfnamefont{M.}~\bibnamefont{{Boylan-Kolchin}}},
  \bibinfo {author} {\bibfnamefont{V.}~\bibnamefont{{Springel}}}, \bibinfo
  {author} {\bibfnamefont{S.~D.~M.}\ \bibnamefont{{White}}},\ and\ \bibinfo
  {author} {\bibfnamefont{A.}~\bibnamefont{{Jenkins}}},\ }%
  \bibfield{journal}{%
  \Doi{10.1111/j.1365-2966.2010.16774.x}{\bibinfo {journal} {Mon. Not. R.
  Astron. Soc.}}\ }%
  \textbf{\bibinfo {volume} {406}},\ \bibinfo {pages} {896} (\bibinfo {month}
  {Jun.}\ \bibinfo {year} {2010}),\
  \Eprint{http://arxiv.org/abs/0911.4484}{arXiv:0911.4484}%
  \bibAnnoteFile{NoStop}{Spin2}%
\bibitem{Peebles}%
  \BibitemOpen
  \bibfield{author}{%
  \bibinfo {author} {\bibfnamefont{P.~J.~E.}\ \bibnamefont{{Peebles}}},\ }%
  \bibfield{journal}{%
  \Doi{10.1086/149876}{\bibinfo {journal} {\apj}}\ }%
  \textbf{\bibinfo {volume} {155}},\ \bibinfo {pages} {393} (\bibinfo {month}
  {Feb.}\ \bibinfo {year} {1969})%
  \bibAnnoteFile{NoStop}{Peebles}%
\bibitem{White}%
  \BibitemOpen
  \bibfield{author}{%
  \bibinfo {author} {\bibfnamefont{S.~D.~M.}\ \bibnamefont{{White}}},\ }%
  \bibfield{journal}{%
  \Doi{10.1086/162573}{\bibinfo {journal} {\apj}}\ }%
  \textbf{\bibinfo {volume} {286}},\ \bibinfo {pages} {38} (\bibinfo {month}
  {Nov.}\ \bibinfo {year} {1984})%
  \bibAnnoteFile{NoStop}{White}%
\bibitem{Dorosh}%
  \BibitemOpen
  \bibfield{author}{%
  \bibinfo {author} {\bibfnamefont{A.~G.}\ \bibnamefont{{Doroshkevich}}},\ }%
  \bibfield{journal}{%
  \bibinfo {journal} {Astrofizika}\ }%
  \textbf{\bibinfo {volume} {6}},\ \bibinfo {pages} {581} (\bibinfo {year}
  {1970})%
  \bibAnnoteFile{NoStop}{Dorosh}%
\bibitem{zeldovich}%
  \BibitemOpen
  \bibfield{author}{%
  \bibinfo {author} {\bibfnamefont{Ya.~B.}\ \bibnamefont{{Zel'dovich}}},\ }%
  \bibfield{journal}{%
  \bibinfo {journal} {Astron. Astrophys.}\ }%
  \textbf{\bibinfo {volume} {5}},\ \bibinfo {pages} {84} (\bibinfo {month}
  {Mar.}\ \bibinfo {year} {1970})%
  \bibAnnoteFile{NoStop}{zeldovich}%
\bibitem{Chandra}%
  \BibitemOpen
  \bibfield{author}{%
  \bibinfo {author} {\bibfnamefont{S.}~\bibnamefont{{Chandrasekhar}}},\ }%
  \bibfield{journal}{%
  \Doi{10.1086/144517}{\bibinfo {journal} {\apj}}\ }%
  \textbf{\bibinfo {volume} {97}},\ \bibinfo {pages} {255} (\bibinfo {month}
  {Mar.}\ \bibinfo {year} {1943})%
  \bibAnnoteFile{NoStop}{Chandra}%
\bibitem{Gerhard}%
  \BibitemOpen
  \bibfield{author}{%
  \bibinfo {author} {\bibfnamefont{O.~E.}\ \bibnamefont{{Gerhard}}}\ and\
  \bibinfo {author} {\bibfnamefont{J.}~\bibnamefont{{Binney}}},\ }%
  \bibfield{journal}{%
  \bibinfo {journal} {Mon. Not. R. Astron. Soc.}\ }%
  \textbf{\bibinfo {volume} {216}},\ \bibinfo {pages} {467} (\bibinfo {month}
  {Sep.}\ \bibinfo {year} {1985})%
  \bibAnnoteFile{NoStop}{Gerhard}%
\bibitem{Merritt}%
  \BibitemOpen
  \bibfield{author}{%
  \bibinfo {author} {\bibfnamefont{D.}~\bibnamefont{{Merritt}}}\ and\ \bibinfo
  {author} {\bibfnamefont{G.~D.}\ \bibnamefont{{Quinlan}}},\ }%
  \bibfield{journal}{%
  \Doi{10.1086/305579}{\bibinfo {journal} {\apj}}\ }%
  \textbf{\bibinfo {volume} {498}},\ \bibinfo {pages} {625} (\bibinfo {month}
  {May}\ \bibinfo {year} {1998}),\
  \Eprint{http://arxiv.org/abs/arXiv:astro-ph/9709106}{arXiv:astro-ph/9709106}%
  \bibAnnoteFile{NoStop}{Merritt}%
\bibitem{Cruz}%
  \BibitemOpen
  \bibfield{author}{%
  \bibinfo {author} {\bibfnamefont{F.}~\bibnamefont{{Cruz}}}, \bibinfo {author}
  {\bibfnamefont{H.}~\bibnamefont{{Vel{\'a}zquez}}},\ and\ \bibinfo {author}
  {\bibfnamefont{H.}~\bibnamefont{{Aceves}}},\ }%
  \bibfield{journal}{%
  \bibinfo {journal} {Revista Mexicana de Astronomia y Astrofisica}\ }%
  \textbf{\bibinfo {volume} {43}},\ \bibinfo {pages} {95} (\bibinfo {month}
  {Apr.}\ \bibinfo {year} {2007})%
  \bibAnnoteFile{NoStop}{Cruz}%
\bibitem{Kalnajs}%
  \BibitemOpen
  \bibfield{author}{%
  \bibinfo {author} {\bibfnamefont{A.~J.}\ \bibnamefont{{Kalnajs}}},\ }%
  in\ \emph{\bibinfo {booktitle} {Dynamics of Disc Galaxies}},\ \bibinfo
  {editor} {edited by\ \bibinfo {editor} {\bibnamefont{{B.~Sundelius}}}}\
  (\bibinfo {year} {1991})\ pp.\ \bibinfo {pages} {323--+}%
  \bibAnnoteFile{NoStop}{Kalnajs}%
\bibitem{Dehnen}%
  \BibitemOpen
  \bibfield{author}{%
  \bibinfo {author} {\bibfnamefont{W.}~\bibnamefont{{Dehnen}}},\ }%
  \bibfield{journal}{%
  \Doi{10.1086/301226}{\bibinfo {journal} {Astron. J.}}\ }%
  \textbf{\bibinfo {volume} {119}},\ \bibinfo {pages} {800} (\bibinfo {month}
  {Feb.}\ \bibinfo {year} {2000}),\
  \Eprint{http://arxiv.org/abs/arXiv:astro-ph/9911161}{arXiv:astro-ph/9911161}%
  \bibAnnoteFile{NoStop}{Dehnen}%
\bibitem{Milos}%
  \BibitemOpen
  \bibfield{author}{%
  \bibinfo {author} {\bibfnamefont{M.}~\bibnamefont{{Milosavljevi{\'c}}}}\ and\
  \bibinfo {author} {\bibfnamefont{D.}~\bibnamefont{{Merritt}}},\ }%
  \bibfield{journal}{%
  \Doi{10.1086/378086}{\bibinfo {journal} {\apj}}\ }%
  \textbf{\bibinfo {volume} {596}},\ \bibinfo {pages} {860} (\bibinfo {month}
  {Oct.}\ \bibinfo {year} {2003}),\
  \Eprint{http://arxiv.org/abs/arXiv:astro-ph/0212459}{arXiv:astro-ph/0212459}%
  \bibAnnoteFile{NoStop}{Milos}%
\bibitem{Sesana}%
  \BibitemOpen
  \bibfield{author}{%
  \bibinfo {author} {\bibfnamefont{A.}~\bibnamefont{{Sesana}}}, \bibinfo
  {author} {\bibfnamefont{F.}~\bibnamefont{{Haardt}}},\ and\ \bibinfo {author}
  {\bibfnamefont{P.}~\bibnamefont{{Madau}}},\ }%
  \bibfield{journal}{%
  \Doi{10.1086/513016}{\bibinfo {journal} {\apj}}\ }%
  \textbf{\bibinfo {volume} {660}},\ \bibinfo {pages} {546} (\bibinfo {month}
  {May}\ \bibinfo {year} {2007}),\
  \Eprint{http://arxiv.org/abs/arXiv:astro-ph/0612265}{arXiv:astro-ph/0612265}%
  \bibAnnoteFile{NoStop}{Sesana}%
\bibitem{NFW}%
  \BibitemOpen
  \bibfield{author}{%
  \bibinfo {author} {\bibfnamefont{J.~F.}\ \bibnamefont{{Navarro}}}, \bibinfo
  {author} {\bibfnamefont{C.~S.}\ \bibnamefont{{Frenk}}},\ and\ \bibinfo
  {author} {\bibfnamefont{S.~D.~M.}\ \bibnamefont{{White}}},\ }%
  \bibfield{journal}{%
  \Doi{10.1086/177173}{\bibinfo {journal} {\apj}}\ }%
  \textbf{\bibinfo {volume} {462}},\ \bibinfo {pages} {563} (\bibinfo {month}
  {May}\ \bibinfo {year} {1996}),\
  \Eprint{http://arxiv.org/abs/arXiv:astro-ph/9508025}{arXiv:astro-ph/9508025}%
  \bibAnnoteFile{NoStop}{NFW}%
\bibitem{Moore}%
  \BibitemOpen
  \bibfield{author}{%
  \bibinfo {author} {\bibfnamefont{B.}~\bibnamefont{{Moore}}}, \bibinfo
  {author} {\bibfnamefont{T.}~\bibnamefont{{Quinn}}}, \bibinfo {author}
  {\bibfnamefont{F.}~\bibnamefont{{Governato}}}, \bibinfo {author}
  {\bibfnamefont{J.}~\bibnamefont{{Stadel}}},\ and\ \bibinfo {author}
  {\bibfnamefont{G.}~\bibnamefont{{Lake}}},\ }%
  \bibfield{journal}{%
  \Doi{10.1046/j.1365-8711.1999.03039.x}{\bibinfo {journal} {Mon. Not. R.
  Astron. Soc.}}\ }%
  \textbf{\bibinfo {volume} {310}},\ \bibinfo {pages} {1147} (\bibinfo {month}
  {Dec.}\ \bibinfo {year} {1999}),\
  \Eprint{http://arxiv.org/abs/arXiv:astro-ph/9903164}{arXiv:astro-ph/9903164}%
  \bibAnnoteFile{NoStop}{Moore}%
\bibitem{Ricotti}%
  \BibitemOpen
  \bibfield{author}{%
  \bibinfo {author} {\bibfnamefont{M.}~\bibnamefont{{Ricotti}}}, \bibinfo
  {author} {\bibfnamefont{A.}~\bibnamefont{{Pontzen}}},\ and\ \bibinfo {author}
  {\bibfnamefont{M.}~\bibnamefont{{Viel}}},\ }%
  \bibfield{journal}{%
  \Doi{10.1086/520113}{\bibinfo {journal} {Astrophys. J. Lett.}}\ }%
  \textbf{\bibinfo {volume} {663}},\ \bibinfo {pages} {L53} (\bibinfo {month}
  {Jul.}\ \bibinfo {year} {2007}),\
  \Eprint{http://arxiv.org/abs/0706.0856}{arXiv:0706.0856}%
  \bibAnnoteFile{NoStop}{Ricotti}%
\bibitem{Cen}%
  \BibitemOpen
  \bibfield{author}{%
  \bibinfo {author} {\bibfnamefont{R.}~\bibnamefont{{Cen}}}, \bibinfo {author}
  {\bibfnamefont{F.}~\bibnamefont{{Dong}}}, \bibinfo {author}
  {\bibfnamefont{P.}~\bibnamefont{{Bode}}},\ and\ \bibinfo {author}
  {\bibfnamefont{J.~P.}\ \bibnamefont{{Ostriker}}},\ }%
  \bibfield{journal}{%
  \bibinfo {journal} {ArXiv Astrophysics e-prints}}%
   (\bibinfo {month} {Mar.}\ \bibinfo {year} {2004}),\
  \Eprint{http://arxiv.org/abs/arXiv:astro-ph/0403352}{arXiv:astro-ph/0403352}%
  \bibAnnoteFile{NoStop}{Cen}%
\bibitem{delpopolo2}%
  \BibitemOpen
  \bibfield{author}{%
  \bibinfo {author} {\bibfnamefont{A.}~\bibnamefont{{Del Popolo}}},\ }%
  \bibfield{journal}{%
  \Doi{10.1111/j.1365-2966.2010.17288.x}{\bibinfo {journal} {Mon. Not. R.
  Astron. Soc.}},\ \bibinfo {pages} {1312}}%
   (\bibinfo {month} {Sep.}\ \bibinfo {year} {2010})%
  \bibAnnoteFile{NoStop}{delpopolo2}%
\bibitem{takada}%
  \BibitemOpen
  \bibfield{author}{%
  \bibinfo {author} {\bibfnamefont{M.}~\bibnamefont{{Takada}}}, \bibinfo
  {author} {\bibfnamefont{E.}~\bibnamefont{{Komatsu}}},\ and\ \bibinfo {author}
  {\bibfnamefont{T.}~\bibnamefont{{Futamase}}},\ }%
  \bibfield{journal}{%
  \Doi{10.1103/PhysRevD.73.083520}{\bibinfo {journal} {\prd}}\ }%
  \textbf{\bibinfo {volume} {73}},\ \bibinfo {pages} {083520} (\bibinfo {month}
  {Apr.}\ \bibinfo {year} {2006}),\
  \Eprint{http://arxiv.org/abs/arXiv:astro-ph/0512374}{arXiv:astro-ph/0512374}%
  \bibAnnoteFile{NoStop}{takada}%
\bibitem{wmap}%
  \BibitemOpen
  \bibfield{author}{%
  \bibinfo {author} {\bibfnamefont{E.}~\bibnamefont{{Komatsu}}}, \bibinfo
  {author} {\bibfnamefont{K.~M.}\ \bibnamefont{{Smith}}}, \bibinfo {author}
  {\bibfnamefont{J.}~\bibnamefont{{Dunkley}}}, \bibinfo {author}
  {\bibfnamefont{C.~L.}\ \bibnamefont{{Bennett}}}, \bibinfo {author}
  {\bibfnamefont{B.}~\bibnamefont{{Gold}}}, \bibinfo {author}
  {\bibfnamefont{G.}~\bibnamefont{{Hinshaw}}}, \bibinfo {author}
  {\bibfnamefont{N.}~\bibnamefont{{Jarosik}}}, \bibinfo {author}
  {\bibfnamefont{D.}~\bibnamefont{{Larson}}}, \bibinfo {author}
  {\bibfnamefont{M.~R.}\ \bibnamefont{{Nolta}}}, \bibinfo {author}
  {\bibfnamefont{L.}~\bibnamefont{{Page}}}, \bibinfo {author}
  {\bibfnamefont{D.~N.}\ \bibnamefont{{Spergel}}}, \bibinfo {author}
  {\bibfnamefont{M.}~\bibnamefont{{Halpern}}}, \bibinfo {author}
  {\bibfnamefont{R.~S.}\ \bibnamefont{{Hill}}}, \bibinfo {author}
  {\bibfnamefont{A.}~\bibnamefont{{Kogut}}}, \bibinfo {author}
  {\bibfnamefont{M.}~\bibnamefont{{Limon}}}, \bibinfo {author}
  {\bibfnamefont{S.~S.}\ \bibnamefont{{Meyer}}}, \bibinfo {author}
  {\bibfnamefont{N.}~\bibnamefont{{Odegard}}}, \bibinfo {author}
  {\bibfnamefont{G.~S.}\ \bibnamefont{{Tucker}}}, \bibinfo {author}
  {\bibfnamefont{J.~L.}\ \bibnamefont{{Weiland}}}, \bibinfo {author}
  {\bibfnamefont{E.}~\bibnamefont{{Wollack}}},\ and\ \bibinfo {author}
  {\bibfnamefont{E.~L.}\ \bibnamefont{{Wright}}},\ }%
  \bibfield{journal}{%
  \bibinfo {journal} {ArXiv e-prints}}%
   (\bibinfo {month} {Jan.}\ \bibinfo {year} {2010}),\
  \Eprint{http://arxiv.org/abs/1001.4538}{arXiv:1001.4538}%
  \bibAnnoteFile{NoStop}{wmap}%
\bibitem{cmbfast}%
  \BibitemOpen
  \bibfield{author}{%
  \bibinfo {author} {\bibfnamefont{U.}~\bibnamefont{{Seljak}}}\ and\ \bibinfo
  {author} {\bibfnamefont{M.}~\bibnamefont{{Zaldarriaga}}},\ }%
  \bibfield{journal}{%
  \Doi{10.1086/177793}{\bibinfo {journal} {\apj}}\ }%
  \textbf{\bibinfo {volume} {469}},\ \bibinfo {pages} {437} (\bibinfo {month}
  {Oct.}\ \bibinfo {year} {1996}),\
  \Eprint{http://arxiv.org/abs/arXiv:astro-ph/9603033}{arXiv:astro-ph/9603033}%
  \bibAnnoteFile{NoStop}{cmbfast}%
\bibitem{bbks}%
  \BibitemOpen
  \bibfield{author}{%
  \bibinfo {author} {\bibfnamefont{J.~M.}\ \bibnamefont{{Bardeen}}}, \bibinfo
  {author} {\bibfnamefont{J.~R.}\ \bibnamefont{{Bond}}}, \bibinfo {author}
  {\bibfnamefont{N.}~\bibnamefont{{Kaiser}}},\ and\ \bibinfo {author}
  {\bibfnamefont{A.~S.}\ \bibnamefont{{Szalay}}},\ }%
  \bibfield{journal}{%
  \Doi{10.1086/164143}{\bibinfo {journal} {\apj}}\ }%
  \textbf{\bibinfo {volume} {304}},\ \bibinfo {pages} {15} (\bibinfo {month}
  {May}\ \bibinfo {year} {1986})%
  \bibAnnoteFile{NoStop}{bbks}%
\bibitem{MaBert}%
  \BibitemOpen
  \bibfield{author}{%
  \bibinfo {author} {\bibfnamefont{C.}~\bibnamefont{{Ma}}}\ and\ \bibinfo
  {author} {\bibfnamefont{E.}~\bibnamefont{{Bertschinger}}},\ }%
  \bibfield{journal}{%
  \Doi{10.1086/421766}{\bibinfo {journal} {\apj}}\ }%
  \textbf{\bibinfo {volume} {612}},\ \bibinfo {pages} {28} (\bibinfo {month}
  {Sep.}\ \bibinfo {year} {2004}),\
  \Eprint{http://arxiv.org/abs/arXiv:astro-ph/0311049}{arXiv:astro-ph/0311049}%
  \bibAnnoteFile{NoStop}{MaBert}%
\bibitem{Davis}%
  \BibitemOpen
  \bibfield{author}{%
  \bibinfo {author} {\bibfnamefont{M.}~\bibnamefont{{Davis}}}, \bibinfo
  {author} {\bibfnamefont{G.}~\bibnamefont{{Efstathiou}}}, \bibinfo {author}
  {\bibfnamefont{C.~S.}\ \bibnamefont{{Frenk}}},\ and\ \bibinfo {author}
  {\bibfnamefont{S.~D.~M.}\ \bibnamefont{{White}}},\ }%
  \bibfield{journal}{%
  \Doi{10.1086/163168}{\bibinfo {journal} {\apj}}\ }%
  \textbf{\bibinfo {volume} {292}},\ \bibinfo {pages} {371} (\bibinfo {month}
  {May}\ \bibinfo {year} {1985})%
  \bibAnnoteFile{NoStop}{Davis}%
\bibitem{Aquarius2}%
  \BibitemOpen
  \bibfield{author}{%
  \bibinfo {author} {\bibfnamefont{V.}~\bibnamefont{{Springel}}}, \bibinfo
  {author} {\bibfnamefont{J.}~\bibnamefont{{Wang}}}, \bibinfo {author}
  {\bibfnamefont{M.}~\bibnamefont{{Vogelsberger}}}, \bibinfo {author}
  {\bibfnamefont{A.}~\bibnamefont{{Ludlow}}}, \bibinfo {author}
  {\bibfnamefont{A.}~\bibnamefont{{Jenkins}}}, \bibinfo {author}
  {\bibfnamefont{A.}~\bibnamefont{{Helmi}}}, \bibinfo {author}
  {\bibfnamefont{J.~F.}\ \bibnamefont{{Navarro}}}, \bibinfo {author}
  {\bibfnamefont{C.~S.}\ \bibnamefont{{Frenk}}},\ and\ \bibinfo {author}
  {\bibfnamefont{S.~D.~M.}\ \bibnamefont{{White}}},\ }%
  \bibfield{journal}{%
  \Doi{10.1111/j.1365-2966.2008.14066.x}{\bibinfo {journal} {Mon. Not. R.
  Astron. Soc.}}\ }%
  \textbf{\bibinfo {volume} {391}},\ \bibinfo {pages} {1685} (\bibinfo {month}
  {Dec.}\ \bibinfo {year} {2008}),\
  \Eprint{http://arxiv.org/abs/0809.0898}{arXiv:0809.0898}%
  \bibAnnoteFile{NoStop}{Aquarius2}%
\bibitem{gan}%
  \BibitemOpen
  \bibfield{author}{%
  \bibinfo {author} {\bibfnamefont{J.}~\bibnamefont{{Gan}}}, \bibinfo {author}
  {\bibfnamefont{X.}~\bibnamefont{{Kang}}}, \bibinfo {author}
  {\bibfnamefont{F.~C.}\ \bibnamefont{{van den Bosch}}},\ and\ \bibinfo
  {author} {\bibfnamefont{J.}~\bibnamefont{{Hou}}},\ }%
  \bibfield{journal}{%
  \bibinfo {journal} {ArXiv e-prints}}%
   (\bibinfo {month} {Jun.}\ \bibinfo {year} {2010}),\
  \Eprint{http://arxiv.org/abs/1007.0023}{arXiv:1007.0023}%
  \bibAnnoteFile{NoStop}{gan}%
\bibitem{zentner}%
  \BibitemOpen
  \bibfield{author}{%
  \bibinfo {author} {\bibfnamefont{A.~R.}\ \bibnamefont{{Zentner}}}, \bibinfo
  {author} {\bibfnamefont{A.~A.}\ \bibnamefont{{Berlind}}}, \bibinfo {author}
  {\bibfnamefont{J.~S.}\ \bibnamefont{{Bullock}}}, \bibinfo {author}
  {\bibfnamefont{A.~V.}\ \bibnamefont{{Kravtsov}}},\ and\ \bibinfo {author}
  {\bibfnamefont{R.~H.}\ \bibnamefont{{Wechsler}}},\ }%
  \bibfield{journal}{%
  \Doi{10.1086/428898}{\bibinfo {journal} {\apj}}\ }%
  \textbf{\bibinfo {volume} {624}},\ \bibinfo {pages} {505} (\bibinfo {month}
  {May}\ \bibinfo {year} {2005}),\
  \Eprint{http://arxiv.org/abs/arXiv:astro-ph/0411586}{arXiv:astro-ph/0411586}%
  \bibAnnoteFile{NoStop}{zentner}%
\end{thebibliography}%

\end{document}